\documentclass[a4paper,11pt]{article}

\usepackage{jheppub} 
                     
\usepackage{amsmath,amssymb,amsthm,amscd,graphicx}
\usepackage{psfrag}
\usepackage[english]{babel}
\usepackage{float}
\input epsf.sty

\usepackage{color}

\theoremstyle{definition}

\newcommand{\CA}{{\cal A}}
\newcommand{\CB}{{\cal B}}

\newcommand{\CN}{{\cal N}}
\newcommand{\CO}{{\cal O}}

\newcommand{\mO}{{\mathsf{O}}}

\def\IR{{\mathbb R}}
\def\IQ{{\mathbb Q}}

\def\IP{{\mathbb P}}

\def\IF{{\mathbb F}}

\newcommand{\re}{{\rm e}}
\newcommand{\ri}{{\rm i}}
\newcommand{\rd}{{\rm d}}

\newcommand{\mx}{{\mathsf{x}}}
\newcommand{\my}{{\mathsf{y}}}

\newcommand{\bea}{\begin{eqnarray}}
\newcommand{\eea}{\end{eqnarray}}
\newcommand{\be}{\begin{equation}}
\newcommand{\ee}{\end{equation}}
\newcommand{\ba}{\begin{aligned}}
\newcommand{\ea}{\end{aligned}}

\newcommand{\sectiono}[1]{\section{#1}\setcounter{equation}{0}}

\newdimen\tableauside\tableauside=1.0ex
\newdimen\tableaurule\tableaurule=0.4pt
\newdimen\tableaustep
\def\phantomhrule#1{\hbox{\vbox to0pt{\hrule height\tableaurule width#1\vss}}}
\def\phantomvrule#1{\vbox{\hbox to0pt{\vrule width\tableaurule height#1\hss}}}
\def\sqr{\vbox{%
  \phantomhrule\tableaustep
  \hbox{\phantomvrule\tableaustep\kern\tableaustep\phantomvrule\tableaustep}%
  \hbox{\vbox{\phantomhrule\tableauside}\kern-\tableaurule}}}
\def\squares#1{\hbox{\count0=#1\noindent\loop\sqr
  \advance\count0 by-1 \ifnum\count0>0\repeat}}
\def\tableau#1{\vcenter{\offinterlineskip
  \tableaustep=\tableauside\advance\tableaustep by-\tableaurule
  \kern\normallineskip\hbox
    {\kern\normallineskip\vbox
      {\gettableau#1 0 }%
     \kern\normallineskip\kern\tableaurule}%
  \kern\normallineskip\kern\tableaurule}}
\def\gettableau#1{\ifnum#1=0\let\next=\null\else
\squares{#1}\let\next=\gettableau\fi\next}

\newcommand{\impart}{\mathrm{Im}}

\newcommand{\figref}[1]{Fig.~\protect\ref{#1}}

\preprint{
{\small{\textsf{NSF-ITP-17-152}}}
}

\title{Non-Perturbative Quantum Mechanics from Non-Perturbative Strings}

\author[a]{Santiago Codesido,}
\affiliation[a]{D\'epartement de Physique Th\'eorique \& Section de Math\'ematiques,\\ Universit\'e de Gen\`eve, Gen\`eve, CH-1211 Switzerland\\}
\emailAdd{santiago.codesido@unige.ch}

\author[a]{Marcos~Mari\~no,}
\emailAdd{marcos.marino@unige.ch}

\author[b,c]{Ricardo~Schiappa\,}
\affiliation[b]{Kavli Institute for Theoretical Physics, University of California,\\ Santa Barbara, CA 93106, United States of America\\}
\affiliation[c]{CAMGSD, Departamento de Matem\'atica, Instituto Superior T\'ecnico,\\ Universidade de Lisboa, 1049-001 Lisboa, Portugal\\}
\emailAdd{schiappa@math.tecnico.ulisboa.pt}

\abstract{This work develops a new method to calculate non-perturbative corrections in one-dimensional Quantum Mechanics, based on trans-series solutions to the refined holomorphic anomaly equations of topological string theory. The method can be applied to traditional spectral problems governed by the Schr\"odinger equation, where it both reproduces and extends the results of well-established approaches, such as the exact WKB method. It can be also applied to spectral problems based on the quantization of mirror curves, where it leads to new results on the trans-series structure of the spectrum. Various examples are discussed, including the modified Mathieu equation, the double-well potential, and the quantum mirror curves of local $\IP^2$ and local $\IF_0$. In all these examples, it is verified in detail that the trans-series obtained with this new method correctly predict the large-order behavior of the corresponding perturbative sectors.}


\arxivnumber{1712.02603}

\begin{document}
\maketitle

\vfill

\eject

\allowdisplaybreaks


\section{Introduction}

It has been known for some time that certain string theories may be encoded in simple quantum mechanical models. This has led to a very fruitful interaction between string theory (and its close cousin, supersymmetric gauge theories) and Quantum Mechanics, as illustrated by \cite{ns,mmrev}. It was recently pointed out that this connection is more general than previously thought. More precisely, it was argued in \cite{cm-ha} that the all-orders WKB approximation in generic one-dimensional quantum systems is encoded in the so-called 
refined holomorphic anomaly equations of topological string theory \cite{bcov,hk,kw}. This makes it possible to 
calculate the quantum periods (also known as Voros multipliers \cite{ddpham}) to \textit{all} orders,  
by using both modularity and the direct integration of the holomorphic anomaly equations \cite{hk06,gkmw}. In the case of polynomial potentials, there is a physical reason for the connection between 
the WKB method and the holomorphic anomaly equations \cite{gg-qm,gm-qm}: 
it turns out that these quantum-mechanical models can be engineered in terms of Seiberg--Witten (SW) theories with gauge group $SU(N)$ (where $N$ corresponds 
to the degree of the potential), in the NS background \cite{ns}, and in a particular scaling limit near the Argyres--Douglas point in moduli space. Since the holomorphic anomaly equation governs the 
quantum periods of supersymmetric gauge theories in this background \cite{mirmor,hkpk, huang}, they are inherited by the quantum-mechanical model.

So far, the results in \cite{cm-ha} are purely perturbative: the $\hbar$ expansion in the WKB method is treated as a genus expansion in topological string theory, and obtained, order by order, by using the recursive structure of the holomorphic anomaly. It is now natural to ask whether the connection between quantum mechanical problems and the holomorphic anomaly goes beyond perturbation theory, and can be used to compute non-perturbative effects in Quantum Mechanics. 
In other words, is it possible to extend this connection to quantum mechanical trans-series\footnote{We refer the reader to \cite{mmlargen, abs17} for an introduction to 
resurgence, trans-series, and their asymptotics, alongside a very complete list of references on the subject.} including exponentially small corrections? 
The key ingredient for such an extension would be a generalization of the refined holomorphic anomaly equations to the realm of trans-series. 

The standard anomaly equation governing the conventional topological string has already been generalized non-perturbatively in \cite{cesv1,cesv2}. This has led for example to a very precise determination of the large-order behavior of the genus expansion for topological string theory on toric Calabi--Yau (CY) threefolds. More recently, it has been used in \cite{cms} to provide a semiclassical decoding of a recent proposal for a non-perturbative topological string partition function \cite{ghm}. Our first goal in this paper is to generalize the work of \cite{cesv1,cesv2} to the refined topological string. This leads to a general trans-series solution for the refined topological string free energy. Since we want to make contact with simple quantum mechanical problems, we focus on the so-called Nekrasov--Shatashvili (NS) limit \cite{ns}. The trans-series solution of the refined holomorphic anomaly equations should correspond, in this limit, to the non-perturbative sector of Quantum Mechanics.

In order to test this idea, we look at two different types of problems. The first type consists of conventional quantum mechanical models in one dimension, involving the 
Schr\"odinger operator. Extensive work since the late 1970s has led to a good understanding of the trans-series structure in this type of examples, by using exact 
versions of the WKB method \cite{voros, silverstone, voros-quartic,ddpham}, multi-instanton calculations in Quantum Mechanics \cite{zinn-justin, zjj1,zjj2}, or the uniform 
WKB approximation \cite{alvarez, alvarez-casares,dunne-unsal}. We start our analysis by looking at the modified 
Mathieu equation, which is closely related to the NS limit of the $\Omega$-deformed SW theory \cite{sw,n, ns}. 
The study of the refined holomorphic anomaly in this model is best implemented by using modular forms, as first pointed out in \cite{hk06}. Interestingly, the study of trans-series 
solutions requires an extension of this framework to deal with exponentially small corrections. We propose an extended modular ring which captures the properties of the trans-series, 
and we test our results against the large-order behavior of the non-holomorphic extension of the quantum periods. Based on this analysis, we 
find a universal form for the one-instanton correction at all orders in $\hbar$. We also determine a higher instanton solution of the holomorphic anomaly equations which agrees 
with the result for the trans-series in Quantum Mechanics, in the holomorphic limit. The structure found for the modified Mathieu equation extends straightftorwardly to other canonical 
examples in Quantum Mechanics, like the double well and the cubic potential studied in \cite{cm-ha}. These results indicate that the correspondence between quantum mechanical problems and the refined holomorphic anomaly equations found in \cite{cm-ha} \textit{extends} to the non-perturbative realm. 

The power of a new method should also be measured by its ability to go \textit{beyond} what is already known. Our new non-perturbative method in Quantum Mechanics 
shows its true strength in a second type of problems, namely, the spectral problems associated to quantum mirror curves. These spectral problems 
involve difference, rather than differential, equations, and they 
have been intensively studied in the last few years. In spite of this, very little is known about their trans-series structure. Our method gives a concrete tool to calculate 
non-perturbative trans-series for the all-orders WKB expansion in this problem. This leads again to predictions for the large-order behavior of the WKB series, which we test in detail 
in the case of local $\IP^2$. In addition, we can deduce from our results the full one-instanton trans-series for the eigenvalues of the difference equation. The systematic, 
perturbative expansion of these eigenvalues has been recently studied in \cite{gu-s}, and we use our results to predict and test the asymptotic behavior of this expansion in the case of local $\IF_0$. 

The all-orders WKB method produces an asymptotic expansion in powers of $\hbar^2$, which is sometimes called the quantum volume. In the case of 
quantum mirror curves, the relation to topological strings has led to a plethora of exact results which in particular promote the quantum volume to a true function. 
It is then natural to ask what is the relation between the asymptotic expansion and the 
exact result. We perform this analysis in the case of local $\IP^2$ and we find that, as in previous related examples \cite{gmz,cms}, the asymptotic expansion of the quantum volume is Borel summable but its Borel resummation does not agree with the exact result. This opens the way for a ``semiclassical decoding'' of the exact quantum volume in terms of a full trans-series, 
which we leave for future work.   

This paper is organized as follows. In Section \ref{sec2} we review the connection between the all-orders WKB method and the refined holomorphic anomaly equations, 
and we study trans-series solutions to these equations in the NS limit. Sections \ref{sec3}, \ref{sec4}, \ref{sec5} and \ref{sec6} are devoted to detailed discussions of examples. In Section \ref{sec3} 
we consider the modified Mathieu equation. We present the trans-series solution to the corresponding holomorphic anomaly equations in terms of 
modular forms. We test this solution against the large-order behavior of the quantum free energies, and we also check that it reproduces known results in Quantum Mechanics. Section \ref{sec4} does a similar analysis for another important example in Quantum Mechanics: the double-well potential. Sections \ref{sec5} and \ref{sec6} are devoted to examples 
based on quantum mirror curves: the case of local $\IP^2$ and the case of local $\IF_0$, respectively. In the first case, we do a comparison between the Borel resummation 
of the quantum free energies and the exact answer obtained from topological string theory. In the second case, we use our results on trans-series to predict the 
large-order behavior of the perturbative series for the energy levels worked out systematically in \cite{gu-s}. We conclude in Section \ref{sec7} and list some open problems. 
Two Appendices contain supplementary material on the topics discussed in the paper. In the first Appendix we present the formulation of 
the refined holomorphic anomaly equations of \cite{kw,hw} in terms of a master equation, while in the second Appendix we present a large-order test of the trans-series obtained 
in Quantum Mechanics for the modified Mathieu equation.

\sectiono{WKB, trans-series and the refined holomorphic anomaly}

\label{sec2}
\subsection{The all-orders WKB method}
\label{sec2wkb}
 In this paper we will be interested in spectral problems in one dimension. One 
 general strategy to attack these problems is to use the all-orders WKB method \cite{dunham} (see for example \cite{bender-wkb, gp2} for clear presentations). 
 In this method, one defines a formal power series in $\hbar^2$ 
 which we will call the {\it quantum volume}, whose 
 classical limit is the volume in phase space defined by a maximal energy $E$. Let us review how this quantum volume is defined in the case of Schr\"odinger operators. We start with the 
 Schr\"odinger equation for a one-dimensional particle in a potential $V(x)$, 
 \be
 \label{schrodinger}
\hbar^2 \psi''(x) +p^2(x) \psi(x)=0, \qquad p(x) ={\sqrt{2(E-V(x))}}, 
\ee
where we have set the mass $m=1$. Let us consider the standard WKB ansatz for the wavefunction, 
\be
\label{yansatz}
\psi(x)=\exp \left\{ {\ri \over \hbar} \int^x Q(x') \rd x' \right\},
\ee
which leads to a Riccati equation for $Q(x)$, 
\be
\label{riccati}
 Q^2(x)-\ri \hbar\frac{\rd Q}{\rd x}(x)=p^2(x). 
\end{equation} 
We now solve for the function $Q(x)$ as a power series in $\hbar$:
\be
\label{yps}
Q(x)=\sum_{k=0}^\infty Q_k(x)\hbar^k. 
\ee
 If we split this formal power series into even and odd powers of $\hbar$, 
 \be
 \label{yyp}
 Q(x) = Q_{\rm odd} (x) + P(x),
\ee
where 
\be
\label{px}
P(x) =\sum_{n \ge 0} P_{n}(x) \hbar^{2n},
\ee
we find that $Q_{\rm odd}(x)$ is a total derivative, 
\be
Q_{\rm odd}(x)={\ri \hbar \over 2} {P'(x) \over P(x)}={\ri \hbar \over 2} {\rd \over \rd x} \log  P(x), 
\ee
and the wavefunction (\ref{yansatz}) can be written as 
\be
\psi(x) ={1\over {\sqrt{ P(x)}}} \exp \left\{ {\ri \over \hbar} \int^x P(x') \rd x' \right\}.
\ee

Let us now consider the Riemann surface $\Sigma$ defined by 
\be
\label{alg-curve}
y^2= p^2(x). 
\ee
We will restrict ourselves to curves of genus one. The turning points of the WKB problem are the points where $p^2(x)=0$, and correspond to the branch points of the curve (\ref{alg-curve}). 
For a curve of genus one there are only two independent one-cycles 
$A$ and $B$ encircling turning points. The $A$-cycle corresponds to an allowed region for the 
classical motion, while the $B$-cycle corresponds to a forbidden region. The period of the one-form $y(x) \rd x$ along the $A$-cycle gives the volume of the classically allowed region in phase space, 
\be
{\rm vol}_0(E)= {1\over 2} \oint_A y (x) \rd x. 
\ee
By using the formal power series in (\ref{px}), we define the {\it quantum volume} as a formal power series in $\hbar^2$, 
\be
\label{ao-vol}
{\rm vol}_{\rm p}(E)=\sum_{n \ge 0} \hbar^{2n} {\rm vol}_n (E), \qquad {\rm vol}_n (E)= {1\over 2} \oint_A P_n(x) \rd x. 
\ee
The all-orders WKB quantization condition is then 
\be
\label{pqc}
{\rm vol}_{\rm p}(E)= 2 \pi \hbar \left( m+{1\over2} \right), \qquad m=0,1,2, \cdots. 
\ee
One important question in Quantum Mechanics (asked e.g. in \cite{bpv}) is whether there is a well-defined function of $E$ and $\hbar$ whose 
asymptotic expansion in $\hbar$, at fixed $E$, coincides with (\ref{ao-vol}). In other words, is there a non-perturbative definition of the quantum volume? 
Surprisingly, such a definition is not known in general. For the Schr\"odinger equation with polynomial potentials, Voros has constructed 
exact quantum volumes for states of definite parity, defined as fixed points of a functional recursion (see e.g. \cite{voros-zeta,voros1997exact}). In the context of integrable systems, the quantum volume 
can be obtained from the so-called Yang--Yang function. The Yang--Yang function can be explicitly written down in some cases via the 
relation to supersymmetric gauge theory \cite{ns} or to topological string theory \cite{ghm,cgm,wzh,hm,fhm}. 
In all those cases, the quantum volume is a well-defined function and its asymptotic expansion agrees with (\ref{ao-vol}). 

We can use the integrals of the differential $y(x) \rd x$ around the two cycles $A$ and $B$ of the curve (\ref{alg-curve}) to define the {\it classical periods},
\be
t= {1\over 2 \pi} \oint_A y (x) \rd x, \qquad t_D= -\ri \oint_B y(x) \rd x, 
\ee
with appropriate choices of branch cuts for the function $y(x)$. It is useful to introduce the {\it prepotential} or 
{\it classical free energy} $F_0(t)$ by the equation 
\be
\label{prepot}
t_D ={\partial F_0 \over \partial t}. 
\ee
We should regard $t$ as a flat coordinate parametrizing the complex structure of the curve (\ref{alg-curve}). If we now use the quantum-deformed differential $P(x) \rd x$, we promote the classical periods to ``quantum'' periods, 
\be
\label{qps}
t(\hbar)= \hbar \nu =  {1\over 2 \pi} \oint_A P (x) \rd x, \qquad \hbar  {\partial F \over 
\partial t(\hbar)}= {\partial F \over 
\partial \nu}=-\ri  \oint_B P(x) \rd x,  
\ee
where we have introduced the variable $\nu =t(\hbar)/\hbar$. Both quantum periods are defined by formal power series expansions in $\hbar^2$, and the leading-order term of this expansion, which is obtained as $\hbar \rightarrow 0$, gives back the classical periods. Note that the quantum $A$-period is proportional to the 
quantum volume, and the all-orders WKB quantization condition reads
\be
\label{num}
\nu= m+{1\over 2}.
\ee
The quantum $B$-period defines the {\it quantum free energy} as a formal power series in $\hbar^2$, where each coefficient is a function 
of the full quantum period $\nu$, 
\be
\label{qfe}
F(\nu)= \sum_{n\ge0 } F_n(\nu) \hbar^{2n-1}. 
\ee
This quantity is the analogue of the NS free energy in $\CN=2$ supersymmetric gauge theories and topological string theory. 
In writing the quantum free energy and the quantum $A$-period we have already 
made a choice of ``frame'', but as it is well-known from SW theory, there is an infinite number of frames related by duality transformations. We will see in the examples below that, in some cases, 
it is convenient to use different frames to perform the analysis. 

The other type of spectral problems that we want to address in this paper are obtained by quantization of mirror curves to toric CY threefolds. In the genus-one case, mirror 
curves can be writen as
\be
\label{mcurve}
\CO(\re^x, \re^y) + \kappa=0, 
\ee
where $\CO(\re^x, \re^y)$ is a polynomial in $\re^x$, $\re^y$. As explained in e.g. \cite{ghm}, we quantize the function 
$\CO(\re^x, \re^y)$ by promoting $x$, $y$ to canonically conjugate Heisenberg operators $\mx$ and $\my$ on $L^2(\IR)$, satisfying the commutation relation 
\be
[\mx, \my] = \ri \hbar. 
\ee
Ordering ambiguities are solved by Weyl's prescription. In this way we obtain a spectral problem of the form 
\be
\label{ope}
\mO (\re^\mx, \re^\my) |\psi \rangle= -\kappa |\psi \rangle, 
\ee
where $|\psi\rangle$ belongs to the domain of the operator $\mO(\re^\mx, \re^\my)$ inside $L^2(\IR)$. 
When working in the $x$ representation for the wavefunctions, $\re^\my$ acts as a difference operator, and the spectral problem (\ref{ope}) can be written as a difference equation. It has been proved in \cite{kama,lst} that, in many cases, the operators $\mO(\re^\mx, \re^\my)$ have a discrete spectrum (more precisely, their inverses are trace class operators in $L^2(\IR)$). 
On the other hand, difference equations can also be solved with the WKB method \cite{dingle}, by using an ansatz of the form (\ref{yansatz}). The leading order term is given by 
\be
Q(x)= y(x)+ \CO(\hbar), 
\ee
where $y(x)$ is the (multivalued) function defined by (\ref{mcurve}). As shown in \cite{mirmor, acdkv}, this WKB analysis makes 
it possible to define quantum periods, similarly to what we have discussed in the context of the Schr\"odinger equation. 
One of the quantum periods defines the quantum volume, as in (\ref{ao-vol}), and this in turn makes it 
possible to write down a perturbative quantization condition of the form (\ref{pqc}). One can also define a quantum free energy, as in (\ref{qps}). 
It was argued in \cite{mirmor, acdkv}, and further tested in \cite{huangNS,hkrs}, that the quantum free energy obtained with the WKB method agrees with the NS limit of the refined topological string 
free energy for the corresponding CY manifold. 

The refined topological string free energy satisfies a generalized or refined set of holomorphic anomaly equations \cite{kw,hk} which extend the 
original construction in \cite{bcov}. In particular, in the case of the difference equations (\ref{ope}), the quantum free energy defined by the WKB method satisfies the 
NS limit of the equations in \cite{kw,hk}. In \cite{cm-ha}, evidence was given that the quantum free energy of many one-dimensional Schr\"odinger problems \textit{also} satisfies these equations, even when the spectral problem is not related to any known supersymmetric gauge theory or topological string theory. It was then conjectured in \cite{cm-ha} that the connection 
between the WKB method and the holomorphic anomaly equations should be valid for general one-dimensional problems\footnote{Some evidence for this conjecture 
has been already found for curves of genus two in \cite{fkn}.}. This conjectural connection, if true, provides a \textit{unified} framework to study spectral problems coming from Schr\"odinger operators and from quantum mirror curves. We will now review in some detail the refined holomorphic anomaly equations and we will study their trans-series solution. 

\subsection{The refined holomorphic anomaly}

\label{sec2rha}

We consider the B-model refined topological string on a local CY manifold described by a Riemann surface $\Sigma$. 
 This could be of the form (\ref{alg-curve}), as arising in ordinary Quantum Mechanics, or a mirror curve defined by an equation 
 like (\ref{ope}). The moduli space of complex structures on the CY manifold is a special K\"ahler manifold with metric $G_{k \bar m}$. In the case of a local CY manifold 
 built upon a Riemann surface $\Sigma$, the metric is related to the period matrix $\tau$ of $\Sigma$ by  
\be
G_{i \bar j}= \frac{2 \pi \ri}{\beta} \left( \tau- \overline \tau \right)_{ij},
\ee
where $\beta$ is a real normalization constant.

We will denote the corresponding 
 covariant derivatives by $D_i$. The prepotential $F_0$ of this special K\"ahler manifold is precisely the function of the moduli defined by (\ref{prepot}). The Yukawa couplings 
 $C_{ijk}$ are defined by  
\be
\label{yukawa}
C_{ijk}={\partial^3 F_0 \over \partial t_i \partial t_j \partial t_k},
\ee
and we can fix $\beta$ so that
\begin{equation}
	 G_{i \bar j} = {\partial^2 F_0 \over \partial t_i \partial t_j} + \text{h.c.}
\end{equation}
An important quantity entering into the formalism is
\be
\overline{C}_{\bar k}^{\ lm}= G^{l \bar p} G^{m \bar n} \overline{C}_{\bar  p \bar n \bar k}. 
\ee
 The basic quantities in refined topological string theory are the perturbative free energies $F^{(g_1, g_2)}(t_i)$, with $g_1, g_2 \ge 0$ (see for example \cite{n,ikv,hk,ckk} for more details). 
 The total free energy is a function of two parameters, $\epsilon_{1,2}$, and it is defined by the asymptotic expansion 
\be
\label{pRTopSt}
F_{\rm ref} (t_i; \epsilon_1, \epsilon_2)  \sim  \sum_{g_1=0}^{+\infty} \sum_{g_2=0}^{+\infty} \left( \epsilon_1 + \epsilon_2 \right)^{2g_1} \left( \epsilon_1 \epsilon_2 \right)^{g_2-1} F^{(g_1,g_2)}(t_i).
\ee
There are two important limits of this quantity. The first one corresponds to $\epsilon_1=-\epsilon_2=g_s$. In this limit, only the perturbative free energies with $g_1=0$ contribute, and we recover 
the standard topological string with string coupling $g_s$. The second limit is the so-called NS limit \cite{ns}, in which we take $\epsilon_1=0$. More precisely, we define the quantum or NS free energy by the limit
\be
\label{NS-limit}
F (t_i, \hbar)= \lim_{\epsilon_1 \to 0} \epsilon_1\, F (t_i; \epsilon_1, \epsilon_2 = \hbar ).
\ee
It has the asymptotic expansion
\be
\label{pNSTopSt}
F(t_i, \hbar) \sim \sum_{n=0}^{+\infty} F^{(n,0)}(t_i) \hbar^{2n-1}, 
\ee
and we shall denote for simplicity
\be
F^{(n,0)}(t_i)= F_n(t_i).  
\ee

The refined topological string energies can be computed with many different techniques: instanton calculus \cite{n}, 
the refined topological vertex \cite{akmv,ikv}, BPS invariants \cite{ckk,noNew}, and, in the case of the NS limit, the WKB method 
mentioned above \cite{mirmor,acdkv}. Another powerful technique is based on the refined holomorphic anomaly equations. These equations exploit the fact that the 
refined free energies can be promoted to non-holomorphic functions of both the moduli $t_i$ and their 
complex conjugates $\bar t_i$, $F^{(g_1, g_2)}(t_i, \bar t_i)$. The refined holomorphic anomaly equations 
govern the anti-holomorphic dependence of these functions, and they read
\be
\label{rha}
{\partial F^{(g_1, g_2)} \over \partial \bar t_k}= {1\over 2 } \overline{C}_{\bar k}^{\ lm} \left( D_l D_m F^{(g_1, g_2-1)} 
+ \sum_{0<r_1+r_2< g_1+g_2} D_l F^{(r_1, r_2)} D_m F^{(g_1-r_1, g_2-r_2)}\right). 
\ee
These equations are valid for $g_1+g_2\ge 2$. In the standard topological string limit, with $g_1=0$, one recovers the BCOV holomorphic anomaly equations \cite{bcov}. 
 In the NS limit, the first term in the r.h.s. of (\ref{rha}) drops out, and we obtain the simplified equations
\be
\label{rha-ns}
{\partial F_n \over \partial \bar t_k}= {1\over 2 } \overline{C}_{\bar k}^{\ lm}  \sum_{r=1}^{n-1} D_l F_r D_m F_{n-r},  \qquad n\ge 2. 
\ee

In this paper we will focus on the case in which $\Sigma$ has genus one, so that the moduli space of complex structures has dimension one. It can be parametrized 
by the elliptic modulus $\tau$, which is related to the prepotential by
\be
\label{tauf0}
\tau= {\beta \over 2 \pi \ri} {\partial^2 F_0 \over \partial t^2}.
\ee
In this case, the anti-holomorphic dependence 
of the free energies can be encoded in a single function, usually called the {\it propagator} $S$, defined by 
\be
\partial_{\bar{t}} S = {\bar{C}_{\bar{t}}}^{\ t t }.
\label{propAndYuk}
\ee
Here, we have denoted the single index by $t$, which refers to the modulus of the CY. 
By using the propagator, we can write the refined holomorphic anomaly equations in the case of curves of genus one as
\be
\label{pRHAE}
{ \partial F^{(g_1,g_2)}  \over \partial S}= \frac{1}{2} \left( D_t^2 F^{(g_1,g_2-1)} +\sum_{0<r_1+r_2< g_1+g_2} D_t F^{(r_1, r_2)} D_t F^{(g_1-r_1, g_2-r_2)}\right),
\ee
and their NS limit (\ref{rha-ns}) as 
	\begin{equation}
		\frac{\partial F_n}{\partial S} = \frac{1}{2} \sum_{r=1}^{n-1} D_t F_r \: D_t F_{n-r}, \qquad n\ge 2. 
		\label{has-recursion}
	\end{equation}
The 
equations (\ref{rha-ns}) have to be supplemented with an explicit expression for $F_1$. It turns out that, in known examples \cite{hk,kw},
\be
\label{f1}
F_1= -{1\over 24} \log \Delta, 
\ee
where $\Delta$ is essentially the discriminant of the curve $\Sigma$ (it can contain additional functions of the moduli). 
Using (\ref{f1}) as the initial condition, as well as the special geometry of the 
moduli space, the holomorphic anomaly equations (\ref{rha-ns}) determine the functions $F_n$ 
recursively, up to a purely holomorphic dependence on the moduli which is usually called the {\it holomorphic ambiguity}. 

When using the refined holomorphic 
anomaly to compute the quantum free energies, it is important to take into account an important subtlety. The argument $t$ of the functions 
$F_n(t)$ is, as it should, the full quantum period $t(\hbar)$ appearing in (\ref{qps}). The anomaly equations typically give the 
$F_n$s as functions of the complex modulus of the curve---parametrized by the elliptic modulus $\tau$ or by the complex parameter appearing in the 
equation of the curve---which we will denote by $z$. However, the relation between $z$ (or $\tau$) and $t(\hbar)$ is the one determined 
by the {\it classical} period \cite{kw,hk}. 

As first noted in \cite{hk06,gkmw}, in the case of curves of genus one, 
it is very useful, both conceptually and computationally, to 
re-express the anomaly equations in the language of modular forms. This leads to convenient parametrizations of the 
free energies and their holomorphic ambiguities, and to a fast symbolic (computational) implementation of the recursion. When the Riemann surface has 
genus one, there is a single Yukawa coupling, which we will denote by 
\be
Y=C_{ttt}. 
\ee
To satisfy (\ref{propAndYuk}), the propagator $S$ can be written as the non-holomorphic modular form \cite{hk06}
\be
\label{propa-e2}
S=-{\beta \over 12}\widehat E_2(\tau, \bar \tau), 
\ee
where $\widehat E_2(\tau, \bar \tau)$ is defined by
\be
\widehat E_2(\tau, \bar \tau)= E_2(\tau)- {3 \over \pi {\rm Im}\, \tau}, 
\ee
and $E_2(\tau)$ is the weight-two Eisenstein series. It is also very useful to introduce the {\it Maass derivative} acting on 
(almost-holomorphic) modular forms of weight $k$,
\be
\label{maass}
D_\tau=\frac{1}{2\pi\ri}\frac{\rd}{\rd\tau}-\frac{k}{4\pi\text{Im} \tau}.
\ee
The refined holomorphic anomaly equations read in this case, in the NS limit, 
\be
\label{ha-modular}
\frac{\partial F_n}{\partial\widehat{E}_2}=-{\beta^3 \over 24} 
Y^2\sum_{r=1}^{n-1}D_{\tau}F_r D_\tau F_{n-r},\qquad n \ge 2. 
\ee
This equation (\ref{ha-modular}) is solved by an expression of the form, 
\begin{equation}
		F_n = \sum_{r=0}^{2n-3} f_{n,r} \widehat{E}_2^r
		\label{gens-ha}
\end{equation}
	where the coefficients $f_{n,r}$ are holomorphic in $\tau$. Since $F_n$ has modular weight zero for $n\ge 1$, the coefficients $f_{n,r}$ have weight $-2r$. The coefficient $f_{n,0}$ is the holomorphic ambiguity. As first explained in \cite{hk06}, one can determine the holomorphic ambiguity by first finding an appropriate parametrization 
	in terms of modular forms, which depends on the particular curve under consideration. The holomorphic ambiguity is written in this way as an unknown linear combination of known modular forms. To determine the coefficients in this linear combination, one imposes boundary conditions at special loci in the moduli space of the curve. 
	This method has been used in e.g. \cite{gkmw, hkpk,ghkk}. We will see concrete implementations of this procedure in the examples of this paper.

\subsection{Trans-series solutions of the refined holomorphic anomaly equations} 

As set-up originally in \cite{bcov}, the holomorphic anomaly equations are inherently perturbative. Due to 
the recursive nature of these equations, it is natural to ask 
whether they can have trans-series solutions which capture non-perturbative effects. This was answered in 
the affirmative in \cite{cesv1}, and further developed and exemplified in \cite{cesv2, c15, csv16, cms}. 
In this section we will extend some aspects of \cite{cesv1,cesv2} to the refined case, focusing on the NS limit which is 
relevant for quantum mechanical problems. 

The first step in constructing the trans-series solution is to obtain a ``master equation'' for the total free energy (\ref{NS-limit}). Such an equation is given by 
\be
\label{masterNSF}
\frac{\partial F}{\partial S} - \frac{1}{2} \hbar \left( D_t F \right)^2 = \frac{1}{\hbar}\, W - U\, D_t F, 
\ee
where 
\be
\ba
U &= D_t F_0, \\
W &= W_0 + \hbar^2\, W_1, 
\ea
\ee
and 
\be
\ba
W_0 &= \frac{\partial F_0} {\partial S} + \frac{1}{2} \left( D_t F_0 \right)^2, \\
W_1 &= \frac{\partial F_1}{\partial S}. 
\ea
\ee
It is easy to see that this reproduces the recursion (\ref{has-recursion}). A more general master equation can be written away from the NS limit, which should provide the starting 
point for an analysis of general trans-series in the refined topological string. We present this master equation in Appendix \ref{ap-master}. 

We now postulate a trans-series \textit{ansatz} for the solution of this master equation, 
rather than the perturbative series \eqref{pNSTopSt}. The simplest {\it ansatz} is a one-parameter trans-series with an infinite number of exponentially small corrections, 
as the one already used in \cite{cesv1,cesv2} for the standard anomaly equation of \cite{bcov}. We will write 
\be
\label{trans-series-NSTopSt}
F (\sigma; \hbar) = \sum_{n=0}^{\infty} \sigma^n\, \re^{-n \CA/\hbar}\, F^{(n)} (\hbar).
\ee
In this equation, $\CA$ is the instanton action, $\sigma$ is a trans-series parameter, and $F^{(n)}(\hbar)$ denotes the perturbative expansion in $\hbar$ around the $n$-th instanton. All multi-instanton sectors are themselves asymptotic series (as the perturbative series \eqref{pNSTopSt} already was)
\be
\label{multi-inst-NSTopSt}
F^{(n)}(\hbar) \sim \sum_{k=0}^{+\infty} \hbar^{k+b^{(n)}} F_k^{(n)},
\ee
\noindent
where $b^{(n)}$ is a ``characteristic exponent'' or ``starting genus''. 

We now proceed as in \cite{cesv1}, i.e. we insert the trans-series \eqref{trans-series-NSTopSt} into the master equation \eqref{masterNSF}. 
One recovers the \textit{perturbative} NS-limit holomorphic anomaly equation \eqref{has-recursion} for the \textit{perturbative} coefficients in \eqref{pNSTopSt},
 alongside a new \textit{non-perturbative} extension for the \textit{non-perturbative} coefficients in \eqref{multi-inst-NSTopSt}. At first non-trivial non-perturbative order one obtains
\be
\label{hol-act}
\partial_S \CA = 0,
\ee
\noindent
i.e. the instanton action is \textit{holomorphic} (since the anti-holomorphic dependence is all contained in the propagator). This is in complete analogy with what happens in the standard topological string case \cite{cesv1}. Note, however, that this condition only occurs if the combination of starting genera 
\be
B_{nm} = b^{(n-m)} - \left( b^{(n)} - b^{(m)} \right)
\ee
 is strictly positive\footnote{Otherwise there will be extra (non-trivial) constraints.}, $B_{nm} > 0$ (this already occurred in the standard topological-string case; see \cite{cesv1} for further details). We shall proceed under this assumption. The holomorphicity condition (\ref{hol-act}) puts little restrictions on the actual value of the action. However, one expects that it is a {\it period} of the CY manifold, as it was postulated in \cite{dmp-np} based on previous insights \cite{bpv,kazakov-kostov,ps09}. By explicitly using holomorphicity of the instanton action, the remaining terms give us the non-perturbative extension of the NS-limit holomorphic anomaly equations. We find,
\be
\label{npNSHAE}
\ba
& \partial_S F_k^{(n)} = - \sum_{i=1}^k \mathcal{D}_i^{(n)} F_{k-i}^{(n)}  \\
&
+ \frac{1}{2} \sum_{m=1}^{n-1}\, \sum_{i=0}^{k-B_{nm}} \left( \partial_t F_{i}^{(m)} - m \left( \partial_t A \right) F_{i+1}^{(m)} \right) \left( \partial_t F_{k-B_{nm}-i-1}^{(n-m)} - \left( n-m \right) \left( \partial_t A \right) F_{k-B_{nm}-i}^{(n-m)} \right)  \\
&
+ \frac{1}{2} \sum_{m=1}^{n-1} \left( \partial_t F_{k-B_{nm}}^{(n-m)} - \left( n-m \right) \left( \partial_t A \right) F_{k-B_{nm}+1}^{(n-m)} \right) \left( - m \left( \partial_t A \right) F_0^{(m)} \right).
\ea
\ee
\noindent
In these equations, the $\mathcal{D}_i^{(n)}$ are operators defined as follows:
\be
\label{npNSHAE-D2h}
\ba
\mathcal{D}^{(n)}_{2i-1} &= n\, \partial_t A\, \partial_t F^{(0)}_i, \qquad i = 1, 2, 3, \cdots, \\
\mathcal{D}^{(n)}_{2i} &= -\partial_t F^{(0)}_i\, D_t, \qquad i = 1, 2, 3, \cdots.
\ea
\ee
Note that the operator in the first line is a multiplicative operator involving no derivatives. The equations (\ref{npNSHAE}) become very simple for the first instanton sector. We find, 
\be
\label{10eq}
\partial_S F_0^{(1)}=0 
\ee
and 
\be
\partial_S F_k^{(1)}= - \sum_{i=1}^k \mathcal{D}_i^{(1)} F_{k-i}^{(1)}, \qquad k \ge 1. 
\ee
For example, for $k=1,2$ we obtain, 
\be
\label{112eq}
\ba
\partial_S F_1^{(1)}&=- \partial_t A \partial_t F^{(0)}_1 F_0^{(1)}, \\
\partial_S F^{(1)}_2&= - \partial_t A \: 
		\partial_t F^{(0)}_1 \: F_1^{(1)} + \partial_t F_1^{(0)} \: \partial_t F_0^{(1)}.
		\ea
		\ee

There is a slightly simpler master equation which we will also use in the following. Let us define
	\begin{equation}
	\label{tildef}
		\widetilde F =\hbar F-F^{(0)}_0.
	\end{equation}
	Then, the holomorphic anomaly equation can be written as
	\begin{equation}
		\partial_S \widetilde{F} -\frac{1}{2}\left(D_t \widetilde{F}\right)^2=0.
		\label{tildef-master}
	\end{equation}
which can be also solved with a trans-series ansatz, as we will see in our examples. 	

One of the signposts of resurgence is that higher instanton corrections ``resurge'' in the large-order behavior of 
lower instanton series (see e.g. \cite{mmlargen,abs17}). In the case we are considering here, the perturbative series is the series of 
NS free energies (\ref{pNSTopSt}). We expect the large-order behavior of this series to be controlled by the first instanton series $F^{(1)}$. 
More precisely, we expect the leading double-factorial behavior  
\begin{equation}
		F^{(0)}_n \sim  \: \sum_{k=0}^\infty \mu_k \:  \frac{\Gamma\left(2n-b-k\right)}{\mathcal{A}^{2n-b-k}}, \qquad n \gg 1, 
		\label{lo-as}
	\end{equation}
	where we have denoted 
	\be
	\label{b-bone}
	b= b^{(1)}+1,
	\ee
	 $\mathcal{A}$ is the smallest action in absolute value, and the coefficients $\mu_n$ are given by the loop corrections in the one-instanton sector, 
	\begin{equation}
		\mu_k = {\Sigma \over 2 \pi \ri} F^{(1)}_k.
		\label{lo-pred}
	\end{equation}
In this equation, $\Sigma$ is a Stokes parameter which has to be determined in each problem.

 In the rest of this paper, we will study the trans-series solution of the NS limit of the refined holomorphic anomaly in various examples.

\sectiono{Examples in Quantum Mechanics: the modified Mathieu equation}
\label{sec3}

We now apply the formalism developed in the previous section to various examples. The first one is the modified Mathieu equation, which has been studied in detail 
in the recent literature (see e.g. \cite{he-miao, basar-dunne, kpt, ashok, bdu-quantum,pt}), due to its connection to SW theory and its quantum deformation \cite{n,ns}. 

\subsection{WKB analysis and the refined holomorphic anomaly}

The modified Mathieu equation describes a one-dimensional, quantum mechanical particle in a $\cosh(x)$ potential. The classical 
Hamiltonian is 
\be
\label{mat-hamil}
H(p,x)= p^2+ 2 \cosh(x), 
\ee
and the corresponding Schr\"odinger equation is 
\be
\label{mmathieu}
\left( -\hbar^2 {\rd^2 \over \rd x^2} + 2  \cosh(x) - E  \right)\psi(x)=0. 
\ee
The spectral problem associated to the modified Mathieu equation leads to an infinite number of discrete energy levels, labelled by an integer $n=0, 1, 2, \cdots$. These energy levels can be computed 
by using the all-orders WKB method. For a given energy $E$, the turning points of the motion are given by $\pm x_+$, where
\be
x_+=\cosh^{-1}\left({E \over 2}\right). 
\ee
The classical volume of phase space is given by the integral, 
\be
{\rm vol}_0(E)=4\int_0^{x_+} \rd x \, \sqrt{E-2 \cosh x}=8\sqrt{E+2} \left[ {\bf K} \left( \frac{E-2}{E+2} \right)
-{\bf E} \left( \frac{E-2}{E+2} \right) \right], 
\ee
where ${\bf K}(m)$, ${\bf E}(m)$ denote the complete elliptic integrals of the first and the second kind, as a function of the squared modulus $m=k^2$. The all-orders WKB method 
gives an asymptotic expansion for the quantum volume, of the form (\ref{ao-vol}). It is possible to calculate the very first orders of this expansion by using conventional techniques. 
One finds, for example, at the next-to-leading order, 
\be
{\rm vol}_1(E)= \frac{2   {\bf K}\left(\frac{2 E}{E+2}-1\right)-E\, {\bf E}\left(\frac{2 E}{E+2  }-1\right)}{6 (E-2) \sqrt{2+E}}.
\ee
As explained in Section \ref{sec2wkb}, the WKB method makes it possible to define quantum periods, as well as a quantum free energy. The quantum volume gives the quantum $A$-period. There is a quantum $B$-period associated to a cycle which goes around the imaginary axis, and is given by 
\be
\label{qA}
a(E, \hbar) =  {1\over  2 \pi \ri} \int_{-\pi \ri }^{\pi \ri } \rd x \, \left(  \sqrt{E-2 \cosh x} + \sum_{n=1}^\infty \hbar^{2n} P_{n}(x) \right). 
\ee
The classical limit of this period will be denoted by $a(E)$, and it is given by 
\be
\label{lmm}
a(E)= {2 {\sqrt{E}} \over \pi} \sqrt{ 1+ {2  \over E}}\, {\bf E} \left( {4 \over 2 +E} \right). 
\ee
The classical prepotential is defined by, 
\be
\label{c-prep}
{\partial F_0 \over \partial a} = {\rm vol}_0(E), 
\ee
where $a=a(E)$ is the classical limit of the $B$-period. Note that, in this problem, we define the prepotential and the quantum free energy in terms of the 
$A$-period. This is because we are choosing here a particular frame, which we will call the ``electric'' frame. There is a dual, ``magnetic'' frame, obtained by performing an $S$ transformation 
which exchanges the $A$ and the $B$ periods, and which is more useful to analyze the spectral problem, as we shall see. 

The prepotential can be computed at large $a$ as 
\be
\label{exp-prep}
F_0(a)= 4 a^2 \log (2 a) -6a^2 -{1\over 2a^2}-{5 \over 64 a^6} -{3\over 64 a^6}+ \CO(a^{-10}).  
\ee
This is, up to a choice of normalization, the prepotential of SW theory \cite{sw}, and the Riemann surface underlying the 
Hamiltonian (\ref{mat-hamil}) is equivalent to the SW curve 
\be
y^2= (x-u)(x^2-1), 
\ee
where $u$, the complex modulus of the curve, is related to the energy in (\ref{mmathieu}) by 
\be
E=2u. 
\ee
This relation can be regarded as a consequence of the connection between SW theory and classical integrable systems 
\cite{mar-war,rus}, since (\ref{mat-hamil}) is the only non-trivial Hamiltonian of the classical 
$N=2$ Toda lattice. 
The classical prepotential $F_0(a)$ can be promoted to a full quantum free energy by the equation 
\be
\label{qvol-mat}
\hbar {\partial F \over \partial a} = {\rm vol}_{\rm p} (E),
\ee
where $a$ is now the full quantum period in (\ref{qA}). The resulting quantity, 
\be
F(a)=\sum_{n \ge 0} F_n(a) \hbar^{2n-1},
\ee
is the NS limit of the Nekrasov free energy for $SU(2)$, $\CN=2$ supersymmetric Yang--Mills theory 
\cite{ns, mirmor} (or, more precisely, its asymptotic expansion in powers of $\hbar^2$). Let us now introduce the ``magnetic'' quantum period 
\be
\label{addef}
a_D(\hbar)= -{\hbar \over 4 \pi} {\partial  F\over \partial a}.
\ee
The corresponding ``magnetic'' or dual quantum free energy 
\be
\label{fd-series}
F_D(a_D, \hbar)=\sum_{n \ge 0} F_{D,n}(a_D) \hbar^{2n-1}
\ee
is defined by  
\be
\label{dualqfe}
\hbar {\partial F_D( a_D, \hbar) \over \partial a_D(\hbar)}= 4 \pi a(\hbar). 
\ee
Then, the all-orders WKB quantization condition can be written as
\be
\label{adnu}
a_D(\hbar)= - {\hbar \nu \over 2}.
\ee
From this quantization condition one can derive in particular the perturbative expansion of the energy $E=E(\nu)$ as a function of the quantum number $\nu$, by simply expanding the quantum 
period $a_D(\hbar)$ around $u=1$, order by order in $\hbar$, and solving for $u$ as a function of $\hbar$ and $\nu$. One finds in this way, 
\be
\ba
{u(\nu, \hbar)-1 \over \hbar}&= \nu +{1\over 64} \left(4 \nu ^2 +1 \right) \hbar -{\nu \over 1024} \left( 4 \nu ^2+3 \right) \hbar ^2\\
&+\frac{\left(80 \nu ^4+136 \nu
   ^2+9\right) \hbar ^3}{131072}+\frac{\left(-528 \nu ^5-1640 \nu ^3-405 \nu \right) \hbar ^4}{4194304}+{\CO}\left(\hbar ^5\right)
   \ea
   \ee
This is precisely what one obtains in the standard perturbative analysis of the modified Mathieu equation, by using for example the BenderWu package \cite{bwpack}. 

In this quantum-mechanical problem, the connection to topological string theory and its geometric engineering limit indicates that the $F_n (a)$ can be computed by 
using the NS limit of the refined holomorphic anomaly equation. As noted in (\ref{f1}), the first correction to the classical prepotential is given by 
\be
\label{f1sw}
F_1(a)= -{1\over 24} \log \, \Delta, \qquad \Delta=u^2-1. 
\ee
In order to obtain efficiently the higher order corrections $F_n(a)$, with $n\ge 2$, we rewrite the holomorphic
 anomaly equations in terms of modular forms, as we explained in Section \ref{sec2rha}. The relevant modular forms are the ones associated to 
 the SW curve, and discussed e.g. in \cite{hk06}. They form a ring with generators
\be
\label{mod-gens-sw}
\ba
			K_2(\tau) &= \vartheta_3^4(q) + \vartheta_4^4(q),\\
			L_2(\tau) &= \vartheta_2^4(q), \\
			\widehat{E}_2(\tau) &= E_2(\tau) -\frac{3}{\pi \: \impart \tau},
		\ea
		\ee
where $\vartheta_i(q)$ are the Jacobi elliptic functions, and $E_2(\tau)$ is the second Eisenstein series. These generators have modular weight $2$. Their argument is 
	\begin{equation}
		q = \re^{\ri \pi \tau}, 
	\end{equation}
	where $\tau$ is the elliptic modulus 
	\begin{equation}
		\tau = \ri \frac{\mathbf{K}\left(\frac{2}{u+1}\right)}{\mathbf{K}\left(\frac{u-1}{u+1}\right)}
		\label{eq:geom:tauOfU}
	\end{equation}
and is related to the prepotential by 
\be
\label{tau-f0}
{\partial^2 F_0 \over \partial a^2}= -4 \pi \ri \tau, 
\ee
so that the constant $\beta$ in (\ref{tauf0}) is $\beta=-1/2$, and
\be
S= {1 \over 24} \widehat{E}_2. 
\ee
We note that the Maass derivative (\ref{maass}) acts on the above generators as 
\be
\label{algebra}
\ba
D_\tau K_2 &= \frac{1}{6} \widehat{E}_2 K_2+\frac{1}{12}\left( 3 L_2^2-K_2^2\right) ,\\
			D_\tau L_2 &= \frac{1}{6} \widehat{E}_2 L_2 +\frac{1}{6} K_2 L_2 ,\\
			D_\tau \widehat{E}_2 &= \frac{1}{12} \widehat{E}_2^2 - \frac{1}{48}\left(K_2^2+3 L_2^2\right). 
			\ea
			\ee

As usual in SW theory, we have to specify the electric-magnetic frame for the calculation of the quantum prepotential. We will reserve the notation $F_n$ for the quantum free energies 
in the electric frame. When $n=0,1$, the free energies in the electric frame are given in (\ref{c-prep}), (\ref{f1sw}). Expressions in the magnetic frame can be obtained, in the language of modular forms, by an $S$-duality transformation
\be
\tau \rightarrow \tau_D =-{1\over \tau}. 
\ee
We also introduce
\be
q_D= \re^{2 \pi \ri \tau_D}. 
\ee
It will be also useful to introduce the quasi-modular forms, 
\be
\label{aad-mod}
\ba
a(\tau) &= \frac{2}{3} \frac{E_2(\tau)+K_2(\tau)}{ \sqrt{L_2(\tau)}}, \\
a_D(\tau_D)& = \frac{2}{3} \frac{E_2(\tau_D) - K_2(\tau_D)-3 L_2(\tau_D)}{\sqrt{2 K_2(\tau_D) - 2 L_2(\tau_D)}}.  
\ea
\ee
The first one corresponds to the electric period $a(E)$, while the second one is related to the $B$-period by 
\be
{\partial F_0 \over \partial a}=- 4 \pi  a_D \left(-1/\tau \right). 
\ee
They have the following expansions in terms of the exponentiated modulus, 
\begin{align}
		a(\tau) &= \frac{1}{\sqrt{q}} \left(\frac{1}{2}+3 q^2-\frac{21 q^4}{2}+33 q^6+\cdots\right), \\
		a_D (\tau_D) &= -16 q_D - 96 q_D^2 - 384 q_D^3 + \cdots.
		\label{eq:geom:aOfq}
	\end{align}
The free energies in the magnetic frame $F_{D,n}$, appearing in (\ref{fd-series}) can be obtained from the $F_n$ by an $S$-transformation, and 
depend on the variable $a_D$. We recall that the holomorphic anomaly equations will give the $F_n$s as functions of the elliptic modulus $\tau$ or its dual $\tau_D$. To obtain them as 
functions of the full quantum periods $a(\hbar)$, $a_D(\hbar)$, we should use the classical equations (\ref{aad-mod}) relating $\tau$ and $\tau_D$ to $a$, $a_D$.

In order to use the anomaly equations (\ref{ha-modular}) we have to specify as well the Yukawa coupling, which is given by 
\be
Y=  \frac{16 \sqrt{L_2}}{K_2^2 - L_2^2}. 
\ee
We also need an appropriate parametrization of the holomorphic ambiguity. In this case, since the relevant ring of modular forms is generated by 
(\ref{mod-gens-sw}), we parametrize the ambiguity by 
\begin{equation}
		f_{n,0} 
			= Y^{2n-2} \sum_{i=0}^{  \left\lfloor \frac{3(n-1)}{2} \right\rfloor  } a_{n,i} \: K_2^{3n-3-2i}L_2^{2i},
	\end{equation}
	where the $a_{n,i}$ are constant numbers. They are fixed by the following boundary conditions for the dual free energies, 
	\begin{equation}
		F_{D,n}(a_D) = \left(\frac{1}{4}\right)^{n-1} \frac{\left(1-2^{1-2n}\right) B_{2n}}{2n(2n-1)(2n-2)}
			\frac{1}{a_D^{2-2n}} + \CO\left(a_D^0\right), \qquad n \ge 2, 
		\label{dual-gap}
	\end{equation}
	where $B_{2n}$ are the Bernouilli numbers. In this way we find, for example, 
	\be
	\ba 
	F_2 &= \frac{Y^2}{4423680} \Big[ 10 \widehat{E}_2 K_2^2+K_2^3-75 K_2 L_2^2 \Big], \\
			F_3 &= \frac{Y^4}{10273695989760} \Big[ -140 \widehat{E}_2^3 K_2^3+\widehat{E}_2^2 \left(840 K_2^4+1260 K_2^2 L_2^2\right)+\\
				&+\widehat{E}_2 \left(21 K_2^5-28287 K_2^3 L_2^2-9450 K_2 L_2^4\right)+\\
				&+769 K_2^6+310500 K_2^2 L_2^4+87012 K_2^4 L_2^2+43875 L_2^6 \Big]. 
				\ea
				\ee
				Higher order free energies can be easily computed recursively. 

\subsection{The free energy trans-series}	

\label{oneinst-sec}
We now proceed to compute the free energy trans-series from the extended holomorphic anomaly equations. We first introduce the covariant derivative w.r.t. the modulus $a$ as
\begin{equation}
	D_a = \frac{D_\tau}{D_\tau a} =\beta Y D_\tau =  -\frac{Y}{2} D_\tau.
	\end{equation}
We look for a trans-series solution of the holomorphic anomaly equations, involving exponentially small quantities of the form $\re^{-\CA/\hbar}$. In order to use the formalism of modular forms, 
we have to take into account that the exponents in these quantities are instanton actions, given by periods, and are not modular invariant. We will now 
introduce a formalism which makes it possible to exploit modularity in spite of this fact. First 
of all, we enlarge the ring of modular objects as follows. Since $E_2$ does not transform as a modular form under an $S$ transformation, we introduce the quantity 
\begin{equation}
		E_2^D = E_2 + \frac{6}{\ri \pi \tau},
		\label{eq:trans:E2Ddef}
	\end{equation}
	so that
	\begin{equation}
			E_2(-1/\tau) = \tau^2 E_2^D(\tau),\qquad 
			E_2^D(-1/\tau) =  \tau^2 E_2(\tau), 
		\end{equation}
		i.e. $E_2$, $E_2^D$ provide a vectorial representation of the $S$ transformation\footnote{Note that $E_2^D$, as defined in (\ref{eq:trans:E2Ddef}), 
		is in principle not invariant under the $T$ transformation $\tau \rightarrow \tau+1$. 
		In order to find the correct results for the large-order behavior, we need however $E_2^D$ to be invariant under this transformation. 
		This can be done by restricting the definition (\ref{eq:trans:E2Ddef}) to the fundamental domain in the upper half plane, and then 
		extend it to the rest of the plane by imposing invariance under $T$.}. In addition, we introduce a degree-counting constant which transforms with 
		weight one under $S$, and which we denote by $\omega_1$. In 
		any evaluation it should be taken to $1$. 
 This leads to an enlarged ring with additional generators $E_2$, $E_2^D$ and $\omega_1$. To define the action of the Maass derivative on these additional 
 generators, we note that $D_\tau$ can be written as 
 \be
 D_\tau = \frac{1}{2\pi \ri} \partial_\tau + \frac{k}{12}\left(\widehat{E}_2-E_2\right), 
 \ee
and we assign a weight $-2$ to $\tau$, so that
\be
		D_\tau\left(\frac{1}{\ri\pi \tau}\right) = -\frac{1}{2} \frac{1}{(\ri\pi \tau)^2} -\frac{1}{6} \frac{\widehat{E}_2-E_2}{\ri \pi \tau}. 
	\end{equation}
 We can then calculate in a straightforward way, 
	\begin{align}
		\begin{split}
			D_\tau E_2 &= \frac{1}{6}\widehat{E}_2 E_2 - \frac{1}{48} \left(K_2^2 +3 L_2^2 +4 E_2^2\right),\\
			D_\tau E_2^D &= \frac{1}{6}\widehat{E}_2 E_2^D - \frac{1}{48} \left(K_2^2 +3 L_2^2 +4 \left(E_2^D\right)^2\right),\\
			D_\tau \omega_1 & = \frac{1}{12}\left(\widehat{E}_2-E_2\right) \omega_1.
		\end{split}
	\label{ext-alg}
	\end{align}
In addition, we postulate the following transformation of $\omega_1$ under $S$-duality, 
\begin{equation}
 \omega_1(-1/\tau)=\tau  \frac{\pi}{6 \ri} \frac{E_2^D - E_2}{\omega_1}.
		\label{tmMap}
	\end{equation}
 Both the Maass derivative of $\omega_1$ and its $S$ transformation lead to a formalism with very useful properties. Let us introduce 
 the actions, 
	\begin{equation}
		\begin{split}
			\mathcal{A}_A(\tau) &=  4\pi a \: \frac{1}{\omega_1}= \frac{8\pi}{3} \frac{E_2 + K_2}{\omega_1 \sqrt{L_2}}, \\
			\mathcal{A}_B(\tau) &=  \ri \,  \omega_1\, \partial_a F_0= \frac{16 \ri\: \omega_1}{\sqrt{L_2}} \frac{E_2^D+K_2}{E_2-E_2^D}, 
		\end{split}
		\label{aab}
	\end{equation}
which now transform with weight zero. As a first check of the usefulness of $\omega_1$, we first note that the derivatives of these actions w.r.t. $a$ are precisely what we expect from (\ref{tau-f0}), namely
	\begin{equation}
	\ba
			D_a \mathcal{A}_A &= -\frac{1}{2} Y  D_\tau \mathcal{A}_A = \frac{4\pi}{\omega_1},\\
		D_a \mathcal{A}_B &= \frac{24 \ri}{E_2-E_2^D} \: \omega_1 = 4 \pi  \tau\,  \omega_1.
		\ea
				\label{daAB}
	\end{equation}
In addition, the introduction of $\omega_1$, together with its $S$ transformation rule in (\ref{tmMap}) 
makes it possible to work out the dual expansion of the actions. For example, after an $S$ transformation, we find, 
\be
			\mathcal{A}_A = 8 \ri \sqrt{2} \frac{  K_2 (\tau_D)+3 L_2(\tau_D) - 2 E^D_2(\tau_D)}{\left[ E_2(\tau_D)-E^D_2(\tau_D)\right] \sqrt{L_2(\tau_D)-K_2(\tau_D)}}\, \omega_1, 
	\end{equation}
	which after setting $\omega_1=1$ leads to the following dual expansion, 
	\be
\mathcal{A}_A = 16+ 4 a_D \left(\log \left(\frac{a_D}{16}\right)-1\right)+
			\frac{3 a_D^2}{4}+\frac{5 a_D^3}{32}+\frac{55 a_D^4}{1024}+\CO(a_D^5). 
			\ee
Similarly, one finds 
\be
\mathcal{A}_B=-4 \pi \ri a_D. 
\ee

Now that we have an appropriate, enlarged modular formalism, let us look for a trans-series solution of the NS holomorphic anomaly equations. 
We will use the master equation (\ref{tildef-master}) for the modified quantum free energy (\ref{tildef}), and we 
consider the trans-series ansatz 
	\begin{equation}
		\widetilde{F} = \widetilde{F}^{(0)} + \widetilde{F}^{(1)}+ \cdots
			\label{1inst-a}
		\ee
Here, $\widetilde{F}^{(1)}$ is the one-instanton correction
\be
 \qquad  \widetilde F^{(1)}= \hbar^{b} f^{(1)} \re^{-G/\hbar}, 
 \label{1inst-ex}
 \ee
where $f^{(1)}$ is a constant and 
\be
\re^{-G/\hbar}= \re^{-\CA/\hbar} \sum_{k\ge 0} F_k^{(1)} \hbar^k, 
\ee
i.e. $G$ is the exponentiated instanton action together with all quantum corrections around the one-instanton configuration. As we shall see, it is possible to 
determine $G$ in a single strike. By plugging the ansatz (\ref{1inst-a}), (\ref{1inst-ex}) in (\ref{tildef-master}), we find
	\begin{equation}
		\partial_S G - D_a\widetilde{F}^{(0)} D_a G = 0.
			\label{geq}\ee
Our goal is now to solve this equation for $G$. We know that $G= \CA +\cdots$. Let us then consider the ansatz
	\begin{equation}
		G = \mathcal{A} + \Big(  \left(S- \mathcal{S_A}\right) D_a \mathcal{A} \Big) D_a \widetilde{F}^{(0)},
		\label{Gansatz}
	\end{equation}
	where $\mathcal{S_A}$ has weight two and is holomorphic
	\begin{equation}
	\label{sa-hol}
		 \partial_S \mathcal{S_A} = 0.
	\end{equation}
Our first observation is that $\partial_{\widehat{E}_2}$ and $D_a$ (or $D_\tau$) do not commute in general: by looking at the algebra of extended generators (\ref{algebra}) and (\ref{ext-alg}), we find that, 
when acting on an object of weight $k$, 
	\begin{equation}
			\partial_{\widehat{E}_2} D_\tau =\frac{k}{12}  + D_\tau \partial_{\widehat{E}_2}.
		\label{eq:mg:dSdTauCommutator}
	\end{equation}
This can be also written as 
	\begin{equation}
		\partial_S  D_a = D_a  \partial_S - k\: Y.
		\label{eq:trans:dSdAcommutator}
	\end{equation}
We deduce that $\partial_S$ and $D_a$ commute when acting on objects of zero weight. We have introduced $\omega_1$ precisely so that actions have zero weight, therefore
	\begin{equation}
	\label{acomm}
		\partial_S \left(D_a \mathcal{A}\right) = D_a \left(\partial_S \mathcal{A}\right) = 0, 
	\end{equation}
	since actions are holomorphic, therefore independent of the propagator, as established in (\ref{hol-act}) (in this context, holomorphy of $\CA$ 
	means that it does not involve the generator $\widehat{E}_2$ of 
	the extended ring). $\partial_S$ and $D_a$ also commute on the weight zero object $\widetilde{F}^{(0)}$. We then calculate 
	\begin{equation}
		\begin{split}
			\partial_S G & = D_a \mathcal{A} \: D_a \widetilde{F}^{(0)} + \Big( \left(S- \mathcal{S_A}\right) D_a \mathcal{A} \Big) D_{aa} \widetilde{F}^{(0)} D_a \widetilde{F}^{(0)},\\
			D_a G &= D_a \mathcal{A} + D_a \Big(  \left(S- \mathcal{S_A}  \right) D_a \mathcal{A} \Big) D_a \widetilde{F}^{(0)} + \Big(\left(S- \mathcal{S_A}\right) D_a \mathcal{A} \Big) D_{aa} \widetilde{F}^{(0)},
		\end{split}
	\end{equation}
and we conclude that (\ref{geq}) is verified if
	\begin{equation}
		D_a \Big(\left(S- \mathcal{S_A}\right) D_a \mathcal{A} \Big) = 0.
		\label{eq:trans:holPropEquation}
	\end{equation}
This equation relates $\mathcal{S_A}$ and $\mathcal{A}$. In fact, when the action $\CA$ is one of the actions in (\ref{aab}), we can solve for $\mathcal{S_A}$ as follows. 
As a consequence of the algebra (\ref{ext-alg}), one has
\begin{equation}
		\begin{split}
			D_a \Big( D_a\mathcal{A}_A \cdot\left(\widehat{E}_2 - E_2\right)\Big) &= D_a \left(4\pi\frac{\widehat{E}_2-E_2}{\omega_1}\right) = 0, \\
			D_a \Big( D_a\mathcal{A}_B \cdot \left(\widehat{E}_2 -E_2^D\right) \Big) &= D_a \left(24 \ri \: \omega_1\frac{\widehat{E}_2-E_2^D}{E_2-E_2^D}\right) = 0.
		\end{split}
		\label{eq:trans:secondDerivativeAction}
	\end{equation}
	We conclude that 
	\begin{equation}
		\begin{split}
			S- \mathcal{S_{A,\mathrm{A}}} &= \frac{1}{24} \left(\widehat{E_2}-E_2\right) = \frac{\ri}{4\pi} \frac{1}{\bar{\tau}-\tau },\\
			S-\mathcal{S_{A,\mathrm{B}}} &= \frac{1}{24} \left(\widehat{E_2}-E_2^D\right) = \frac{1}{4\pi \tau} \frac{\ri \bar{\tau}}{\bar{\tau}-\tau}.
		\end{split}
		\label{sss}
	\end{equation}
Finally, upon using (\ref{daAB}), we find:
	\begin{equation}
	\ba
		G_A(\tau,\bar{\tau})\: \omega_1&= 4\pi a(\tau) + \frac{\ri}{\bar{\tau}-\tau} D_a\widetilde{F}^{(0)}(\tau,\bar{\tau}),\\
		G_B(\tau,\bar{\tau})\: \omega_1^{-1}&= 
		\ri \: \partial_a F_0(\tau)
		+\frac{\ri \: \bar{\tau}}{\bar{\tau}-\tau} D_a \widetilde{F}^{(0)}(\tau,\bar{\tau}).
		\label{GABex}
		\ea
	\end{equation}
This gives the full one-instanton correction after exponentiation. Let us note that the functions $G_{A, B}$ have a very non-trivial $\hbar$ expansion. For $G_A$, one finds, 
	\begin{equation}
\ba
			G_A (\tau, \bar \tau) \omega_1&=4\pi  a(\tau)-\frac{\pi  \left(\widehat{E}_2-E_2\right) K_2 \sqrt{L_2}}{36  \left(K_2^2-L_2^2\right)} \hbar^2 \\
			&+
				\frac{\pi  \left(\widehat{E}_2-E_2\right) L_2^{3/2}}{311040  \left(L_2^2-K_2^2\right){}^3}
				 \Big[-20 \widehat{E}_2^2 K_2^2+K_2^4+40 \widehat{E}_2 \left(2 K_2^3+3 K_2 L_2^2\right)\\ & \qquad \qquad -1347 K_2^2 L_2^2-450 L_2^4\Big]\hbar^4+ \CO\left(\hbar^6\right), 
\ea
		\label{eq:trans:gFunctionAperiod}
	\end{equation}
while for $G_B$, one has
\begin{equation}
			\begin{split}
				G_B(\tau, \bar \tau)\omega _1^{-1} &= \ri \: \partial_a F_0 +\frac{\ri K_2 \sqrt{L_2}   \left(\widehat{E}_2-E_2^D\right)}{6 \left(E_2^D-E_2\right) \left(K_2^2-L_2^2\right)} \: \hbar ^2+ \\
				&+\frac{\ri L_2^{3/2}  \left(\widehat{E}_2-E_2^D\right) }{51840 \left(E_2-E_2^D\right) \left(K_2^2-L_2^2\right)^3}
				\Big[K_2^2 \left(20 \widehat{E}_2^2-80 \widehat{E}_2 K_2-K_2^2\right)\\
				& \qquad +3 K_2 L_2^2 \left(449 K_2-40 \widehat{E}_2\right)+450 L_2^4\Big]\:\hbar ^4+\CO\left(\hbar ^6\right).
			\end{split}
			\label{eq:trans:gFunctionBperiod}
		\end{equation}
By expanding the exponential, we can immediately calculate all the coefficients $F_k^{(1)}$, $k \ge 1$. 

We should mention that trans-series for the quantum free energies associated to the Mathieu equation have been computed in \cite{kpt} by using WKB methods. 
They find solutions involving exponentially small corrections in the periods, as we have found above.  

An immediate application of the above calculation is the determination of the trans-series for the quantum volume function, which in this case is given by (\ref{qvol-mat}). 
We can write it as 
	\begin{equation}
		V^{(0)}={\rm vol}_{\rm p}(E)= {\partial F_0 \over \partial a}+ D_a \widetilde F, 
	\end{equation}
	calculated in the electric frame. The one-instanton trans-series associated to this is of the form 
	\be
	\ba
	V^{(1)}&= D_a \left( \re^{-G/\hbar} \right)= - \hbar^{b-1} D_a G \:  \re^{-G/\hbar} \\
	&= -\hbar^{b-1}
		\left(D_a\mathcal{A} + \Big(\left(S- \mathcal{S_A}\right) D_a \mathcal{A} \Big) D_{aa} \widetilde{F}^{(0)}\right) \re^{-G/\hbar}.
		\ea
		\ee
	For instance, for the $B$-period action we find
	\begin{equation}
		V^{(1)} = -\hbar^{b-1} \left( 4\pi \tau  +
		\frac{\ri \bar{\tau}}{\bar{\tau}-\tau}  D_{aa} \widetilde{F}^{(0)}  \right) \: \omega_1
		\exp\left\{ { -\frac{1}{\hbar}\frac{\ri \bar{\tau}}{\bar{\tau}-\tau}  D_{a} \widetilde{F}^{(0)} }\:  \omega_1 \right\}.
	\end{equation}

\subsection{Application: large-order behavior} 
We have now determined the full one-instanton correction to the quantum free energy, 
for arbitrary values of $\tau$ and $\bar \tau$ (which can be taken to be independent variables). This correction depends on a choice of instanton action, which 
corresponds to the $A$ or the $B$ period. Some ingredients in the answer, like the exponent $b$, are still undetermined. In addition, what we have actually shown is that the expressions (\ref{GABex}) solves 
the master equation, but there might be other solutions differing by a holomorphic ambiguity. For all these reasons, it is important to test our results (\ref{GABex}) by using the resurgent connection 
between one-instanton amplitudes and the large-order behavior of the perturbative series, summarized in (\ref{lo-as}) and (\ref{lo-pred}).

\begin{figure}[h]
		\begin{center}
			\includegraphics[width=0.6\textwidth]{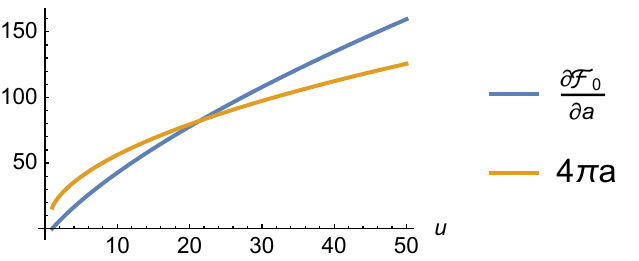}
		\end{center}
		\caption{Absolute value of the instanton actions (\ref{math-act}), as functions of $u$.}
		\label{fig:ab-actions}
	\end{figure}

A numerical study of the large-order behavior of the $F_n^{(0)}$ indicates that $b=2$. In addition, there are two relevant actions, 
\be 
\label{math-act}
\mathcal{A}_A=4\pi a, \qquad 
			\CA_B =   \ri \: \partial_a F^{(0)}_0 ,
	\end{equation}
which are obtained from (\ref{aab}) once we set $\omega_1=1$	. The relative dominance of these actions depends 
on where we are in moduli space, as explained in \cite{dmp-np}. As shown in \figref{fig:ab-actions}, the $A$-period dominates the asymptotics at large $u$, while the $B$-period dominates 
the asymptotics at small $u$, and there is an exchange of dominance as we move from the strong-coupling region of small $u$ to the weak coupling region of large $u$. We then expect that 
the full large-order behavior of the $F_n^{(0)}$ is controlled by the one-instanton amplitudes associated to the $A$ and the $B$ periods, which in turn can be obtained by exponentiating the functions 
appearing in (\ref{GABex}), respectively. The Stokes parameter $\Sigma$ appearing in (\ref{lo-pred}) can be extracted from the boundary behavior in (\ref{dual-gap}), and turns out to be given by 
\be
{\Sigma_A \over 2 \pi \ri}={1\over  \pi^2}, \qquad {\Sigma_B \over 2 \pi \ri}={1\over 2 \pi^2}. 
\ee
Note that the non-holomorphic $F_n^{(0)}$ depends on $q$, $\bar q$ through the modular form
\be
\widehat E_2(q, \bar q)= E_2(q) + {6 \over \log(q \bar q)}, 
\ee
and we can regard $q$, $\bar q$ as independent variables. The trans-series also depends on $q$, $\bar q$, and should control the large-order behavior for arbitrary values of these 
two variables. 

In order to test the predictions (\ref{lo-pred}) for the coefficients $\mu_m$, we can proceed as follows. By using the values of $F_n^{(0)}$ and the predicted values for $\mu_\ell$, $\ell=0, \cdots, m-1$, we consider the sequence
	\be
	\label{num-mum}
	\mu_m^{(n)}= \left(\frac{2n}{\mathcal{A}}\right)^m \left[\frac{F^{(0)}_n \mathcal{A}^{2n-2}}{ \Gamma(2n-2) }
			- \sum_{r=0}^{m-1}\frac{\mu_r\: \mathcal{A}^r}{(2n-2-r)_r} \right], \qquad n=0,1,2, \cdots,
	\end{equation}
	where $(x)_n = x (x+1)\cdots(x+n-1)$ is the Pochhammer symbol. This sequence should converge to $\mu_m$ as $n \rightarrow \infty$. In addition, we can accelerate the convergence of this sequence by using Richardson transforms (in the present context, see for example \cite{msw, abs17}). We will denote by $\mu_m^{(k, \ell)}$ the numerical approximation to $\mu_m$ obtained by taking the first $k$ terms of the sequence $\mu_m^{(n)}$ and performing $\ell$ Richardson transforms. Let us now present some concrete numerical tests of the large-order behavior. Take for example the value, 
\be
q={1\over 10}, 
\ee
which gives $u=1.492...$. This is in the strong-coupling region, and the relevant one-instanton amplitude is associated to the $B$-period. 
The holomorphic limit is obtained when $\bar \tau \rightarrow \ri \infty$, or equivalently when $\bar q \rightarrow 0$. In this limit, and from the explicit value of $G_B$, we find the predictions: 
\begin{equation}
\ba
			\mu_0 &= \frac{1}{2\pi^2},\\
			\mu_1 &= \ri \: 0.00795571145895174009795449858\dots,\\
			\mu_2&=     -0.00062468027457634687835468457\dots,\\
			&\vdots\\
			\mu_5 &= \ri \:  0.00200651427778427678850353789\dots.
		\ea
	\end{equation}
\begin{figure}[h]
		\begin{center}
			\includegraphics[width=0.8\textwidth]{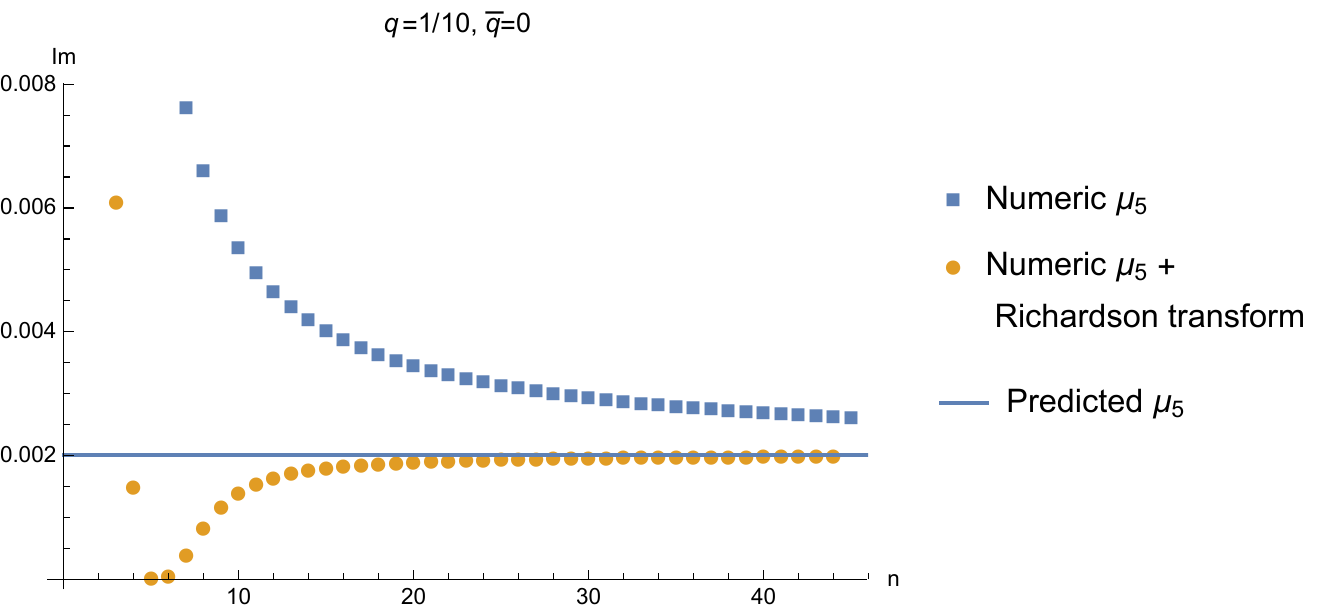}
		\end{center}
		\caption{The sequence (\ref{num-mum}) with $m=5$ for $q=1/10$, $\bar q=0$, up to $n=45$, as well as its first Richardson transform. The horizontal line is the prediction obtained from $G_B$ in 
		(\ref{GABex}).}
		\label{fig-mu5}
	\end{figure}
	The numerical value obtained for $\mu_5$ by using the sequence of the $F_n$ up $n=45$ and with $10$ Richardson transforms is 
	\begin{equation}
		\mu_{5}^{(45,10)} = \ri \: 0.00200651427778\dots,
	\end{equation}
matching $12$ digits with the resurgent prediction of $\mu_5$.  This is a strong test of this \textit{and} the previous coefficients, which were used as input for the numerics. 
In \figref{fig-mu5} we show the sequence (\ref{num-mum}) up to $n=45$, its first Richardson transform, 
and the predicted value for $\mu_5$ from the trans-series. 

Similar tests can be done for more general values of $\bar q$, and in all cases we find 
perfect agreement between the prediction obtained from $G_B$ and the actual large-order behavior. An interesting case occurs when $\bar \tau=0$. This corresponds to the 
free energies in the magnetic frame, $F_{D,n}^{(0)}$, and in the holomorphic limit. In this case, $G_B=\CA_B$ and the asymptotics is trivial, in the sense that 
\be
\mu_0 = {1 \over 2 \pi^2}, \qquad \mu_{n \ge 1}=0. 
\ee
This is precisely what is found in a numerical study of the large-order behavior. We have tested the predictions obtained from our analytic instanton results 
also in the region of large $u$, where the relevant instanton action is associated 
to the $A$-period.

\subsection{Higher-order instanton corrections}
\label{higher-section}
In the previous section we have studied the one-instanton trans-series, but we expect higher instanton corrections. In this section we find a solution to the holomorphic anomaly equations describing 
multi-instanton corrections. First of all, we introduce the following multi-instanton ansatz for the function $\widetilde F$ introduced in (\ref{tildef}):
\begin{equation}
\widetilde{F} = \sum_{m=0}^{\infty} \widetilde{F}^{(m)}, 
\end{equation}
	where each $\widetilde{F}^{(m)}$ has the following structure:
	\be
	\widetilde{F}^{(m)}= \phi^{(m)} \re^{-m G/\hbar}. 
	\ee
We note that 
\be
\phi^{(0)}=\widetilde F^{(0)}=\hbar F^{(0)}- F^{(0)}_0. 
\ee
We have already solved for $\phi^{(1)}$ in Section \ref{oneinst-sec}, namely 
	\be
	\label{tildef1}
	\phi^{(1)}= \hbar^2 f^{(1)}, 
	\ee
	where we used the value of $b=2$ obtained from the large-order analysis. 
After plugging it into the master equation (\ref{tildef-master}), and using (\ref{geq}), we obtain, for $m\ge 1$, 
\be
\ba
			&\partial_S \phi^{(m)} -D_a \phi^{(0)}\cdot D_a \phi^{(m)}  \\
			& = \frac{1}{2}\sum_{r=1}^{m-1} 
			\left(D_a \phi^{(r)} - \frac{r}{\hbar} \phi^{(r)} \cdot D_a G\right)
			\left(D_a \phi^{(m-r)} - \frac{m-r}{\hbar} \phi^{(m-r)} \cdot D_a G\right).
		\ea
		\label{higher-ic}
\ee
	The left hand side of this equation involves the operator that annihilates $G$ in (\ref{geq}), which we will denote by
	\begin{equation}
		W := \partial_S - D_a \phi^{(0)} \cdot D_a.
	\end{equation}
	It has the following properties:
	\begin{enumerate}
		\item $W\left(G\right) = 0$,\:\:  $W\left(D_a \phi^{(0)} \right) =0$,
		\item $W$ is linear,
		\item $W$ is a derivation.
	\end{enumerate}
	In order to solve (\ref{higher-ic}) we have to ``integrate'' with respect to $W$. 
	
	Let us first consider the $m=2$ instanton correction. By using (\ref{tildef1}), we obtain from (\ref{higher-ic}) the equation 
	\begin{equation}
	\label{weq2}
		W \left(\phi^{(2)} \right) = \frac{1}{2} \left(\hbar f^{(1)} \right)^2 \left(D_a G\right)^2.
	\end{equation}
Let us denote by
	\begin{equation}
		T := D_a \mathcal{A} \left(S-\mathcal{S_A}\right)
		\label{Tfunction}
	\end{equation}
	the combination appearing in (\ref{Gansatz}), so that
	\be
	\label{gt}
	G=\mathcal{A} + T \: D_a \phi^{(0)}. 
	\ee
	We recall from (\ref{eq:trans:holPropEquation}) that
	\begin{equation}
	\label{dat0}
		D_a T = 0,
	\end{equation}
	and by using (\ref{sa-hol}) and (\ref{acomm}) we find that 
	\begin{equation}
		\partial_S T = D_a \mathcal{A}. 
	\end{equation}
	Now we apply $W$ on the following weight zero object,
	\begin{equation}
		\begin{split}
			W(T \: D_a G) &= D_a \mathcal{A} \: D_a G + T \: D_a \partial_S G - D_a \phi^{(0)} \: T \: D_{aa} G  \\
			 & = D_a \mathcal{A} \: D_a G + T \: D_{aa}\phi^{(0)} \: D_a G \\
			 & = D_a\left(\mathcal{A} + T \: D_a \phi^{(0)}\right) D_a G = \left(D_a G\right)^2, 
		\end{split}
	\end{equation}
	where we have used (\ref{geq}) and (\ref{dat0}). 
	We conclude that 
	\begin{equation}
		\phi^{(2)} = \phi^{(2)}_0  + \frac{1}{2}\left(\hbar\:f^{(1)}\right)^2 \: T\: D_a G
	\end{equation}
	solves (\ref{weq2}), with $\phi^{(2)}_0$ in the kernel of $W$. The simplest solution compatible with the large-order behavior of the quantum free energies is that $\phi^{(2)}_0=\hbar^2 f^{(2)}$ 
	is a constant, so we find
		\begin{equation}
		\phi^{(2)}/\hbar^2 = f^{(2)}  + \frac{1}{2}\left(f^{(1)}\right)^2 \: T\: D_a G.
		\label{eq:trans:2instSolution}
	\end{equation}

	It is now clear that $T$ always accompanies the $D_a$ derivatives so that the full object keeps the correct weight. Define
	\begin{equation}
		\begin{split}
			\mathcal{W}\left(\cdot\right) &= T^2\:W\left(\cdot\right),\\
			\mathcal{D}\left(\cdot\right) &= T\: D_a \left(\cdot\right).
		\end{split}
		\label{eq:trans:curlyDerivations}
	\end{equation}
	Just like $W$ and $D_a$, they are both linear derivations.
	The recursion for $m \geq 1$ becomes
	\begin{equation}
		\mathcal{W} \phi^{(m)} = \frac{1}{2}\sum_{r=1}^{m-1} 
		\left(\mathcal{D} \phi^{(r)} - \frac{r}{\hbar} \phi^{(r)} \cdot \mathcal{D} G\right)
		\left(\mathcal{D} \phi^{(m-r)} - \frac{m-r}{\hbar} \phi^{(m-r)} \cdot \mathcal{D} G\right).
		\label{eq:trans:instantonRecursionCompact}
	\end{equation}
	Suppose now that $X_0$ has weight zero, so that $\partial_S$ and $D_a$ commute when acting on it. Let us use (\ref{gt}) to write, 
	\begin{equation}
		D_a \phi^{(0)} = \frac{G-\mathcal{A}}{T}.
	\end{equation}
	Then the action of $\mathcal{W}$ can be written in terms of $G$,
	\begin{equation}
		\mathcal{W} X_0 = T^2 \: \partial_S X_0 - \left(G-\mathcal{A}\right) \: \mathcal{D} X_0.
	\end{equation}
	We then have the following commutator,
\begin{equation}
\ba
\mathcal{W} \mathcal{D} X_0 &= T^2 \: \partial_S \left(T\: D_a X_0\right) - \left(G-A\right) \mathcal{D}^2 X_0  \\
			&= T^2 \left(D_a \mathcal{A} \cdot D_a X_0 + T\: D_a\partial_S X_0\right) - \left(G-A\right)\mathcal{D}^2 X_0  \\
			&= \mathcal{D}\mathcal{A}\cdot \mathcal{D} X_0 + \mathcal{D} \left(\mathcal{W}X_0 + \left(G-A\right)\mathcal{D}X_0\right) - \left(G-A\right)\mathcal{D}^2 X_0\\
			&= \mathcal{D}\mathcal{W} X_0 + \mathcal{D}G \cdot \mathcal{D} X_0.
\ea
		\label{WDcom}
	\end{equation}
	This means that $\mathcal{W}$, appearing in (\ref{eq:trans:instantonRecursionCompact}), closes over
	\begin{equation}
		\left<G, \:\mathcal{D} G, \:\mathcal{D}^2 G, \: \mathcal{D}^3 G, \cdots\right>,
	\end{equation}
	and we can build the recursion with these elements.
	Let us look for example at the third-order instanton, with $m=3$. The equation determining $\phi^{(3)}$ is
	\begin{equation}
		\begin{split}
			\mathcal{W} \phi^{(3)} &= 2 \hbar^2  f^{(1)}  f^{(2)} \big(\mathcal{D} G\big)^2 
				+	\hbar^2 \left( f^{(1)}\right)^3 \big(\mathcal{D} G\big)^3 
				- 	\frac{\hbar^3 \left(f^{(1)} \right)^3}{2} \big(\mathcal{D} G\big)  \left(\mathcal{D}^2 G\right).
		\end{split}
	\end{equation}
	The building blocks to solve this equation can be obtained by using (\ref{WDcom}) and $\mathcal{W}G=0$, and we find
	\begin{equation}
		\begin{split}
			\mathcal{W}\left(\frac{1}{3} \mathcal{D}^2G\right) &= \mathcal{D} G \cdot \mathcal{D}^2G, \\
			\mathcal{W}\left(\frac{1}{2} \left(\mathcal{D} G\right)^2\right) &= \left(\mathcal{D} G\right)^3, \\
			\mathcal{W}\Big( \mathcal{D} G\Big) &= \left(\mathcal{D} G\right)^2.
		\end{split}
	\end{equation}
	Therefore,
	\begin{equation}
		\begin{split}
			\phi^{(3)}/\hbar^2 &=  f^{(3)} + 2 f^{(1)}  f^{(2)} \: \mathcal{D} G  + \frac{\left(f^{(1)}\right)^3}{2} \left(\mathcal{D} G\right)^2
				- \frac{\hbar \left(f^{(1)}\right)^3}{6}\mathcal{D}^2 G.
		\end{split}
		\label{f3-sol}
	\end{equation}
	%
Proceeding in this way, it is possible to calculate the $m$-th multi-instanton correction in terms of a set of constants $f^{(1)}, \cdots, f^{(m)}$. In principle these constants should be 
related, since we expect a single trans-series parameter. In the case of the trans-series controlling the large-order behavior of the $F_n$s, for example, the values of 
these constants can be found, in principle, by using explicit large-order results and resurgence relations. 
In the next section we will fix the values of these constants by comparing to results obtained in Quantum Mechanics.  

\subsection{Comparison with previous results in Quantum Mechanics}

In conventional Quantum Mechanics one can find exact quantization conditions for the spectrum by using the 
exact WKB method \cite{voros,voros-quartic,ddpham}, instanton calculus \cite{zinn-justin,zjj1, zjj2}, or the  
uniform WKB approximation \cite{alvarez, alvarez-casares, dunne-unsal}. These quantization conditions are in fact equations defining implicitly a trans-series for the 
quantum period $\nu$ in (\ref{qps}). This leads, by a formal trans-series expansion, to a trans-series for any function of $\nu$, like for example the energy $E=E(\nu)$. 
The exact spectrum is then obtained by applying Borel--\'Ecalle resummation to the resulting trans-series. 

The trans-series obtained from exact quantization conditions are usually based on instanton solutions to the Euclidean 
EOM. However, there are no real instanton solutions for the $\cosh(x)$ potential, and one needs complex instantons \cite{bpv} coming from classical 
trajectories along the imaginary axis in the complex $x$ plane \cite{stone,bdu}, where we have a periodic potential. One way to find the appropriate trans-series for the 
modified Mathieu equation is to start with the $\cos(x)$ potential (i.e., the Mathieu equation). In the $\cos(x)$ potential the quantization condition  
was obtained in \cite{zinn-justin, zjj1, zjj2} by using instanton calculus and derived in \cite{dunne-unsal} by using the uniform WKB method. It reads
\be
\label{sg-qc}
1+\re^{\pm 2\pi \ri \nu}= f_{\rm SG} (\nu)+ 2 \cos \theta {\sqrt{f_{\rm SG}(\nu)}}. 
\ee
Here, $\theta$ is the quasimomentum, and $f_{\rm SG}(\nu)$ can be written as 
\be
f_{\rm SG}(\nu)=  {2 \pi \over \Gamma^2\left(\nu+{1\over 2}\right)} \left( {32 \over \hbar} \right)^{2 \nu} \re^{-A_{\rm SG} (\nu, \hbar)}, 
\ee
where $A_{\rm SG}(\nu, \hbar)$ is a certain regular function of $\nu$, $\hbar$. The $\pm$ sign in (\ref{sg-qc}) corresponds to the choice 
of lateral resummation. To make contact with the modified Mathieu equation, we change $\hbar \rightarrow -\hbar$ in the function $A_{\rm SG}(\nu, \hbar)$, and we eliminate the 
dependence in $\theta$ by taking $\theta=\pi/2$. We end up with the equation, 
\be
\label{nufnu}
 1+\re^{\pm 2\pi \ri \nu}=f(\nu), 
 \ee
 where
\be
\label{fnumathieu}
f(\nu)= {2 \pi \over \Gamma^2\left(\nu+{1\over 2}\right)} \left( {32 \over \hbar} \right)^{2 \nu} \re^{A(\nu, \hbar)}, \qquad 
A(\nu, \hbar)=-A_{\rm SG}\left(\nu, -\hbar \right).
\ee
The function $f(\nu)$ turns out to be related to the derivative of the dual quantum prepotential introduced in (\ref{dualqfe}), as follows:
\be
f(\nu)= \exp\left( {\partial F_{D} (a_D, \hbar) \over \partial a_D(\hbar)}\right), 
\ee
where $a_D(\hbar)$ and $\nu$ are linked by (\ref{adnu}). We recall from (\ref{dual-gap}) that $\partial F_D(a_D, \hbar)/\partial a_D(\hbar)$ has a singular part at $a_D=0$. After exponentiation, the singular part 
gets resummed into the pre-factor of $f(\nu)$ involving the Gamma function. 
$A(\nu, \hbar)$ is then the regular part of this derivative. It has the expansion, 
\be
A(\nu, \hbar)={16 \over \hbar}+ {\hbar \over 16} \left( 3  \nu^2 +{3\over 4} \right) -\left( {\hbar\over 16}\right)^2 \left( 5 \nu^3 + {17 \nu \over 4} \right)+\CO(\hbar^3). 
\ee
The relation (\ref{nufnu}) defines a trans-series for the quantum period $\nu$. To compute it explicitly, we write, as in \cite{alvarez-casares, alvarez},
\be
\label{nu-trans}
\widehat \nu= \nu + \Delta \nu,  \qquad \Delta \nu= \sum_{k \ge 1} \Delta \nu ^{(k)},
\ee
where 
\be
\Delta \nu ^{(k)} \propto \re^{ k A(\nu, \hbar)/\hbar}
\ee
is a $k$-instanton contribution. From the equation (\ref{nufnu}) one finds (we pick the $+$ sign for simplicity)
\be
\label{nu-inst}
\ba
\Delta \nu ^{(1)}&= { \ri \over 2 \pi} f, \\
\Delta \nu ^{(2)}&= {\ri \over 4 \pi} f^2-{1\over 4 \pi^2} f f', \\
\Delta \nu^{(3)}&= \frac{\ri f^3}{6 \pi }-
				\frac{3 f^2 f' }{8 \pi ^2}-
				\frac{\ri f^2 f''}{16 \pi ^3}-
				\frac{\ri f ( f')^2}{8 \pi ^3}, 
\ea
\ee
and so on. In the following, we will rewrite $f(\nu)$ as
	\begin{equation}
		f(\nu)= \exp \left\{- \frac{2}{\hbar} F_D ' \right\}
	\end{equation}
	where we denoted $F_D ' = \partial_\nu F^{(0)}_D\left(\nu, \hbar\right)$. The extra $2/\hbar$ comes from the relation
	between $a_D$ and $\nu$. 
The trans-series for $u(\nu, \hbar)$ (or, equivalently, for the energy) can be obtained by promoting the perturbative relation $u=u(\nu)$ to a trans-series relation, as 
explained in e.g. \cite{alvarez-casares, alvarez}. We obtain, 
\be
\label{transu}
u(\nu + \Delta \nu)= \sum_{n \ge 0} u^{(n)}(\nu). 
\ee
The first three instanton corrections for $u$ read, 
	\begin{equation}
		\begin{split}
		u^{(1)} &=\re^{- 2 F_D '/\hbar}{\ri  u' \over 2 \pi},\\
		u^{(2)} &= \re^{- 4 F_D '/\hbar\:} \left( {\ri u' \over 4 \pi}- {u'' \over 8 \pi^2} + {F_D' \over 2 \pi^2 \hbar} \right), \\
			u^{(3)} &=  \re^{- 6F_D '/\hbar\: }
			\left(\frac{\ri \: F_D'''\: u'}{8 \pi ^3 \hbar}+\frac{\ri \: F_D'' \: u''}{4 \pi ^3 \hbar }-
			\frac{3 \ri \left(F_D''\right)^2 u'}{4 \pi ^3 \hbar ^2}+\frac{3 F_D'' u'}{4 \pi ^2 \hbar }-
			\frac{\ri u'''}{48 \pi ^3}-\frac{u''}{8 \pi ^2}+\frac{\ri u'}{6 \pi } \right),
	\end{split}
		\label{transu-3}
	\end{equation}
where we have denoted $u=u^{(0)}$, $u'=\partial_\nu u$. To verify that (\ref{nufnu}) gives the correct trans-series for the energy, 
we have checked in detail that (\ref{transu}), (\ref{transu-3}) 
lead to the appropriate large-order behavior of the perturbative energy series and of the first 
instanton series; see Appendix \ref{app-mmathieu}. 

Let us now show that (\ref{transu}), (\ref{transu-3}) are compatible with the results obtained from the holomorphic anomaly equation. In the context of the anomaly equation, we obtain 
a trans-series for the quantum free energy, so the comparison is easier to made if there is a functional relation between $u$ and $F$ (or $F_D$) which can be promoted 
to a trans-series relation. In fact, such an equation exists and it is often referred to as ``perturbative/non-perturbative'' or PNP relation\footnote{In spite of their name, PNP relations are 
relations between {\it perturbative} Voros multipliers or quantum periods in one-dimensional quantum systems. They do not carry information about the 
relevant trans-series. This information is encoded in quantization conditions like (\ref{nufnu}).}. PNP relations in one-dimensional quantum systems were first noted by 
G. \'Alvarez and his collaborators in a series of papers \cite{alvarez-casares, alvarez-casares2, alvarezhs, alvarez}, and they
 have been generalized to other models in \cite{dunne-unsal}. In the case of the (modified) Mathieu equation, 
the PNP relation coincides with the extension of the Matone relation \cite{matone} to the quantum NS limit \cite{francisco}, see for example \cite{basar-dunne,bdu-quantum, gorsky}. 
In our notation, the PNP relation reads: 
\begin{equation}
		u(\nu, \hbar) -1 = \frac{\hbar}{32} + \frac{\hbar^3}{8} \: \partial_\hbar \left[ \frac{F_D(\nu,\hbar)}{\hbar} \right]. 
		\label{qMatone}
	\end{equation}
Let us now search for the appropriate trans-series of the quantum free energies which leads, through (\ref{qMatone}), to (\ref{transu}), (\ref{transu-3}). In view of (\ref{dualqfe}), 
the ``classical'' action should be
	\begin{equation}
		\mathcal{A} = - 4\pi a.
	\end{equation}
	Now we need the $G$ function corresponding to this action. Since $f(\nu)$ was written in terms of $F_D$,
	we should write it in the magnetic frame. From
	(\ref{GABex}), we find
	\begin{equation}
		G\left(\tau, \bar{\tau}\right) = -\frac{4\pi a}{\omega_1} - \frac{\ri}{\omega_1 \left( \bar{\tau} -\tau \right)} D_a \widetilde F^{(0)}.
	\end{equation}
	To go to the magnetic frame, we perform an $S$ transformation. Since $\omega_1$
	has weight one, it also transforms, and we obtain
	\begin{equation}
		G\left(\tau_D, \bar{\tau}_D\right) = - \frac{\partial F_{D,0}^{(0)}}{\partial a_D} \omega_1 -
			 \omega_1 \left( -\frac{1}{\tau_D}\right) \frac{\ri \: D_a \widetilde F^{(0)}}{ 1/\tau_D -  1/\bar{\tau}_D  }.
	\end{equation}
	Writing it in the form,
	\begin{equation}
		G = \mathcal{A} + T \: D_a \widetilde{F}^{(0)},
	\end{equation}
	 we have
	\begin{equation}
		\mathcal{A} =  - \frac{\partial F_{D,0}^{(0)}}{\partial a_D} \:\: \omega_1, \qquad 
		T = \frac{\ri \: \omega_1}{1 - \tau_D / \bar{\tau}_D}.
	\end{equation}
	The holomorphic limit in the magnetic frame corresponds to $\bar{\tau}_D \to \ri \infty$. Therefore, in this limit, 
	\begin{equation}
		T = \ri.
	\end{equation}
	The covariant derivative also gets an $\ri$ factor in the magnetic frame (this is due to the fact that, under 
	an $S$ transformation, $a$ goes to $-\ri a_D$). Therefore, in this frame, 
	\begin{equation}
	{\cal D}=T\: D_a  = - \partial_{a_D}=  \frac{2}{\hbar} \partial_\nu,
		\label{eq:trans:covDerivativeMagFrame}
	\end{equation}
where we have used the relation (\ref{adnu}). The $G$ function becomes
	\begin{equation}
		G = -\partial_{a_D} F_{0,D}^{(0)} -
			 \partial_{a_D} \widetilde{F}_{D}^{(0)} = -\hbar \,\partial_{a_D}F_{D}^{(0)}=2\, \partial_\nu F_D. 
		\label{finalG}
	\end{equation}
This is precisely what is needed. Let us now denote by $\widetilde F^{(k)}_D$ the instanton corrections obtained in Section \ref{higher-section} 
in the magnetic frame, and with the above choice (\ref{finalG}) 
for the function $G$. According to (\ref{qMatone}), we have
	\begin{equation}
		u^{(k)} = \frac{\hbar^3}{8} \partial_\hbar \left[ \frac{\widetilde F_D^{(k)} }{\hbar^2} \right], \qquad k \ge 1. 
		\label{ukfk}
	\end{equation}
We have verified that this relation holds true for $k=1, \cdots, 5$ with the following choice of the constants in the holomorphic ambiguity:
	\begin{equation}
		f^{(m)} = \frac{1}{(2m)^2 \: \pi \ri}. 
		\label{fm-mat}
	\end{equation}	
As an example, let us consider the three-instanton free energy. From (\ref{f3-sol}) and (\ref{ukfk}) we get
	\begin{equation}
		\begin{split}
			u^{(3)} & =  \re^{- 6/\hbar \: F_D '}\:
			\left[\frac{8 \left(f^{(1)}\right)^3 F_D''' \: u'}{\hbar }+\frac{16 \left(f^{(1)}\right)^3 F_D''\: u''}{\hbar }-
			\frac{48 \left(f^{(1)}\right)^3 \left(F_D''\right)^2 u'}{\hbar ^2} \right. \\
			& \left.  \qquad \qquad \qquad  -\frac{48 f^{(2)} f^{(1)}\: F_D'' \: u'}{\hbar }-\frac{4 \left(f^{(1)}\right)^3 u'''}{3} +
			8 f^{(2)} f^{(1)} u''-6 f^{(3)} u' \right].
		\end{split}
	\end{equation}
This reproduces precisely the last line in (\ref{transu-3}), once (\ref{fm-mat}) is used. We conclude that our trans-series solution of the holomorphic anomaly equations not 
only leads to the correct large-order behavior of the quantum free energies, but it also reproduces correctly the trans-series obtained from exact quantization conditions in 
Quantum Mechanics. In the next section we will see 
more examples in quantum-mechanical models.

\sectiono{More examples in Quantum Mechanics}
\label{sec4}
\subsection{The double-well potential}

The double-well potential in Quantum Mechanics is given by
	\begin{equation}
		V(x) = \frac{x^2}{2}\left(1+\lambda x\right)^2.
		\label{eq:exactq:DWpotential}
	\end{equation}
	We will set $\lambda=1$. A detailed analysis of the all-orders WKB method in terms of the refined holomorphic anomaly was already presented in \cite{cm-ha}. 
	Here we summarize some of the results. The classical $A$-period 
	$t$ corresponds to the allowed region, while the classical $B$-period $t_D= \partial_t F_0$ 
	corresponds to the tunneling region between the wells. They define together a classical prepotential $F_0(t)$. The modulus $\tau$ 
	of the corresponding elliptic curve satisfies:
	\be
	\label{tau-f0-dw}
\tau={1\over 4 \pi \ri} {\partial^2 F_0 \over \partial t^2}, 
\ee
and the energy is related to $\tau$ by the relationship
	\begin{equation}
		E= \frac{L_2^2}{32 K_2^2}, 
	\end{equation}
	where $K_2$, $L_2$ are the modular forms introduced in (\ref{mod-gens-sw}). From (\ref{tau-f0-dw}) we deduce that $\beta=1/2$, therefore
	\begin{equation}
		S = -\frac{1}{24} \widehat{E}_2.
	\end{equation}	
	The direct integration of the resulting holomorphic anomaly equations, in the NS limit, makes it possible to calculate 
	the functions $F_n$ systematically, as shown in \cite{cm-ha}.

	\begin{figure}[h]
		\centering \includegraphics[width=0.6\textwidth]{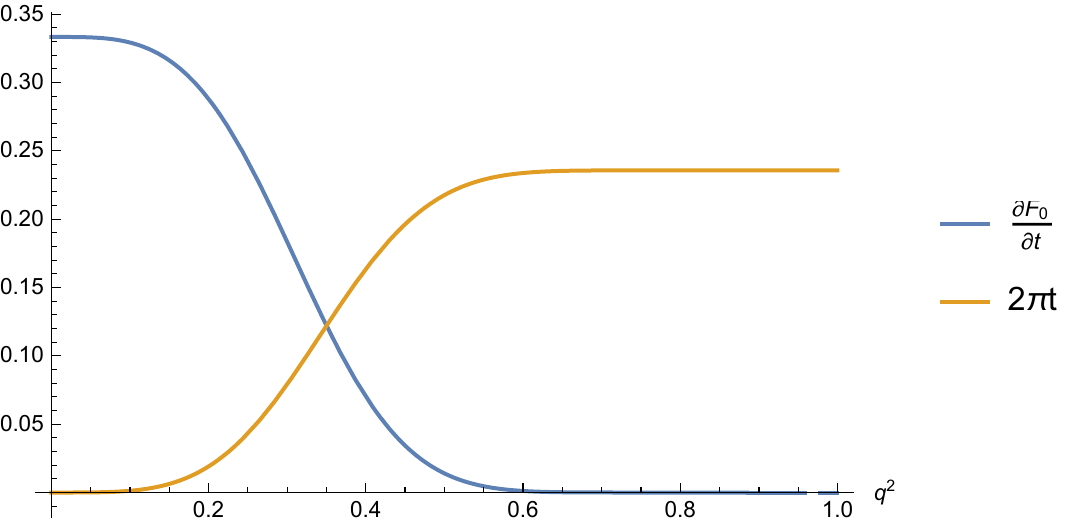}
		\caption{The instanton actions (\ref{dwact}) as a function of $q^2$. }
		\label{fig:exactq:DWactions}
		\end{figure}

We can now use the 	techniques developed in this paper to obtain the one-instanton trans-series. Since quantum-mechanical problems associated to genus-one curves have the same structure, the one-instanton correction is still given by the general solution (\ref{1inst-ex}), where $G$ is given in (\ref{Gansatz}). 
	Equivalently, we can write it as (\ref{gt}), where $T$ is given by (\ref{Tfunction}).  The only ingredient that changes is the parameter $\beta$ 
	appearing in (\ref{propa-e2}), which does not affect the derivation of  (\ref{gt}).
	
Let us now find the trans-series responsible for the large-order behavior of the quantum free energies in the double well. 
	There are two relevant instanton actions, given by the periods
	\begin{equation}
	\label{dwact}
	\ba
		\mathcal{A}_A \:  \cdot\: \omega_1 &= 2\pi\ri\, t ,\\
		\mathcal{A}_B \:/\: \omega_1& = t_D = \partial_t F^{(0)}_0.
		\ea
	\end{equation}
	In \figref{fig:exactq:DWactions} we plot their absolute value as a function of $q^2$, where we clearly see their regions of dominance. 
By using the explicit formulae for $\CA$, we obtain from (\ref{Tfunction}):
	\begin{equation}
		\begin{split}
			T = D_t \mathcal{A} \:\:  (S-\mathcal{S}_A)
			 = D_t \mathcal{A} \left(-\frac{1}{2}\right)\frac{\widehat{E}_2 - E_2^\mathcal{A}}{12}=\left\{\begin{array}{lll}
				\frac{2\pi i}{24} \frac{\widehat{E}_2 - E_2}{\omega_1} & \mathrm{if} & \mathcal{A}=\mathcal{A}_A,\\[2ex]
				\frac{\widehat{E}_2 - E_2^D}{E_2-E_2^D} \: \omega_1 & \mathrm{if} & \mathcal{A}=\mathcal{A}_B.
			\end{array}\right.
		\end{split}
		\label{eq:exactq:DWTfunctions}
	\end{equation}
	The normalization of the trans-series can be determined by the singular behaviour at the conifold points identified in \cite{cm-ha}, which correspond to the energies $E=0$, $E=1/32$. One finds, 
	\begin{equation}
		f^{(1)}_A = \frac{2 \ri}{\pi}, \qquad 	f^{(1)}_B = \frac{\ri}{\pi}.
	\end{equation}
	By plugging now the above results in (\ref{1inst-ex}), we find, for the instanton correction associated to the $\CA_A$ action, 
	\begin{equation}
		\begin{split}
			\hbar^{-2} \widetilde F^{(1)}_A &=\frac{2\ri}{\pi} \re^{-\CA_A/\hbar} \left(1 +
				\frac{\ri \sqrt{2} \pi  \left(\widehat{E}_2-E_2\right) K_2^{3/2} \left(2 K_2^2-3 L_2^2\right)}{9 L_2^2 \omega _1 \left(K_2^2-L_2^2\right)}\hbar \right. \\ 
			&\qquad \qquad \left. 
				-\frac{\pi ^2 \left(\widehat{E}_2-E_2\right){}^2 K_2^3 \left(2 K_2^2-3 L_2^2\right){}^2}{81 L_2^4 \omega _1^2 \left(K_2^2-L_2^2\right){}^2}\hbar^2 + \CO\left(\hbar^3\right)\right) ,
		\end{split}
	\end{equation}
	while for the $\mathcal{A}_B$ action, 
	\begin{equation}
		\begin{split}
			\hbar^{-2} \widetilde F^{(1)}_B &= \frac{\ri}{\pi} \re^{-\CA_B/\hbar} \left(1 + 
			\frac{4 \sqrt{2} K_2^{3/2} \omega _1 \left(\widehat{E}_2-E_2^D\right) \left(2 K_2^2-3 L_2^2\right)}{3 L_2^2 \left(E_2^D-E_2\right)
				\left(L_2^2-K_2^2\right)} \hbar + \right. \\
			&+ \left.
			 \frac{16 K_2^3 \omega _1^2 \left(\widehat{E}_2-E_2^D\right){}^2 \left(2 K_2^2-3 L_2^2\right){}^2}{9 L_2^4 \left(E_2-E_2^D\right){}^2
				\left(K_2^2-L_2^2\right){}^2} \hbar^2 + \CO\left(\hbar^3\right)
			\right).
		\end{split}
	\end{equation}

	\begin{figure}[h]
			\centering \includegraphics[width=0.75\textwidth]{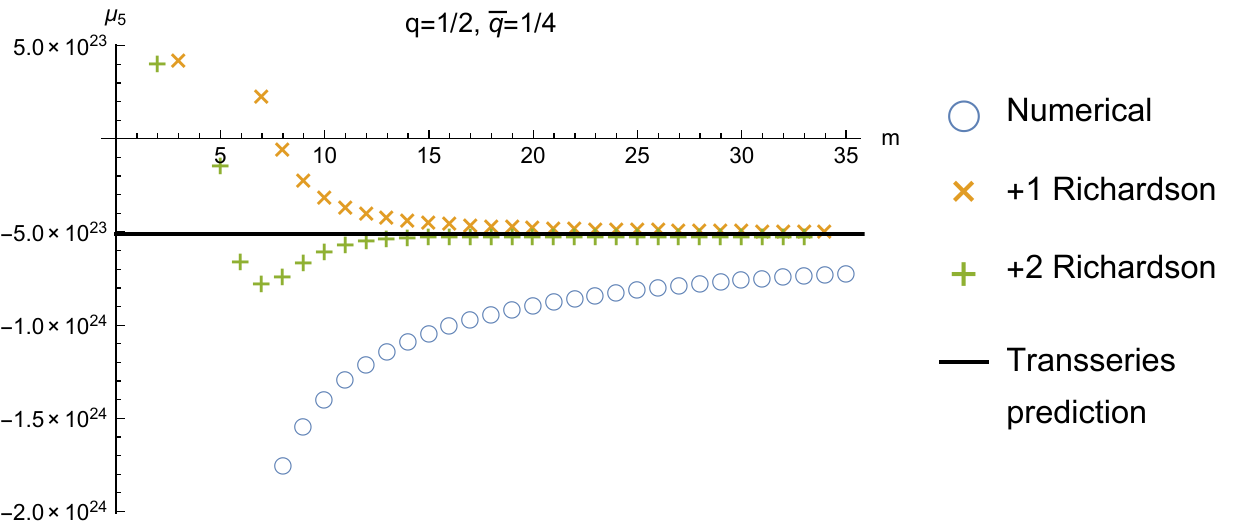}
			\caption{The sequence (\ref{num-mum}) with $m=5$ for the quantum free energies of the double well, at $q=1/2, \, \bar q=1/4$, represented by blue circles (denoted by ``numerical''), 
			as well as its first two Richardson transforms. The horizontal line is the prediction from the trans-series.} 
			\label{fig:exactq:DWmu5}
	\end{figure}
We have tested the above expressions systematically, in different regions of dominance for the instanton actions, and for different 
values of $q$, $\bar q$ (not necessarily complex conjugate values). 
Let us give an example, corresponding to the values
	\begin{equation}
		q = 1/2, \qquad \bar{q} = 1/4.
	\end{equation}
From \figref{fig:exactq:DWactions} we find that these values are in the region where the $B$-period dominates. Using the sequence defined by (\ref{num-mum}), 
	we can compute numerical approximations to the value of $\mu_m$. As before, we denote by $\mu^{(n,r)}_m$ the numerical approximation obtained by taking $n$ values of this sequence, 
	as well as $r$ Richardson transforms. We find, for example, 
	\begin{equation}
		\mu_{5}^{(40,11)} = -5.13138\dots\relax\cdot 10^{23}.
	\end{equation}
	The prediction from the trans-series is
	\begin{equation}
		\left. \frac{ \Sigma }{2\pi \ri}F^{(1)}_5 \right|_{
				\begin{subarray}{l}
					q=1/2\\ \bar{q}=1/4
				\end{subarray}
			} = -5.13138091\dots\relax\cdot 10^{23}. 
			\label{mu5dw}
	\end{equation}
We show the convergence to this value in \figref{fig:exactq:DWmu5}. 

\subsection{Comparison with the exact quantization condition}

As in the case of the modified Mathieu equation, we can compare the trans-series obtained with the refined holomorphic anomaly equations, 
with the trans-series obtained from the exact quantization condition in Quantum Mechanics. The exact quantization condition for the double-well potential 
was first obtained in \cite{zinn-justin} with instanton techniques, and then derived in \cite{ddpham} with the 
exact WKB techniques of Voros--Silverstone \cite{voros,silverstone}\footnote{See however \cite{power} for an approach to the exact energy levels of the 
double-well and other potentials which does not rely on trans-series.}. For us, the most convenient form for this quantization condition is the one derived by G. \'Alvarez in \cite{alvarez} 
by using the uniform WKB method. It reads, 
	\begin{equation}
		1+\re^{2\pi \ri t(\hbar)/\hbar} =\ri  \epsilon \: \re^{-t_D\left( t, \hbar\right)/2\hbar}. 
		\label{dwqc}
	\end{equation}
In this equation, $t(\hbar)=\hbar \nu$ is the quantum $A$-period, $\epsilon$ takes into account the parity of the states, and $t_D(t, \hbar)$ is the quantum $B$-period, 
re-expressed in terms of the quantum $A$-period. From now on we will set $\epsilon=+1$ for simplicity. Note that (\ref{dwqc}) is very similar to the equation (\ref{nufnu}) 
appearing in the context of the modified Mathieu equation, and it can be also used 
to define a trans-series for $\nu$ (or equivalently, the quantum $A$-period $t$), as we did in (\ref{nu-trans}), (\ref{nu-inst}). Any function of $\nu$, like the energy, gets promoted to a trans-series as it 
happened in (\ref{transu}). We find, similarly to (\ref{transu-3}), 
\begin{equation}
		\begin{split}
\widehat E(t,\hbar) &= E(t, \hbar)
				 -\frac{\hbar  \:E '}{2 \pi }
			 \re^{-t_D/2\hbar} + \left( 
				-\frac{\hbar  \:t _D' \: E '}{8 \pi ^2}+\frac{\hbar ^2 \:E ''}{8 \pi ^2}-\frac{\ri \hbar  \:E '}{4 \pi }
			\right) \re^{-t_D/\hbar} + \\
			&+ \left(
				\frac{\hbar ^2 \:t _D' \:E ''}{16 \pi ^3}+\frac{\hbar ^2 \:t _D'' \:E '}{32 \pi ^3}-\frac{3 \hbar  \:t _D'{}^2 \:E '}{64 \pi ^3}-  \frac{3 \ri \hbar  \:t _D' \:E '}{16 \pi ^2}-\frac{\hbar ^3 \:E'''}{48 \pi ^3}+\frac{\ri \hbar ^2 \:E ''}{8 \pi ^2}+\frac{\hbar  \:E '}{6 \pi }
			\right) \re^{-3t_D/2\hbar} \\
			& +\CO\left(\re^{-2t_D/\hbar}\right),
		\end{split}
		\label{eq:exactq:dwEnInstExp}
	\end{equation}
	where we have denoted $t_D'=\partial_t t_D(t,\hbar)$ and $E'=\partial_t E(t,\hbar)$. 
	
	Let us now show how the result above can be reproduced by using the trans-series for the free energy. The 
	only ingredient we need is the PNP relation for the double well obtained in \cite{alvarez}, which we write in the form
	\begin{equation}
		\frac{\partial E}{\partial t} = \xi \: \partial_\lambda \frac{\partial F^{(0)} }{\partial t},
	\end{equation}
	where $\xi$ is an appropriate constant. Suppose now that we choose an instanton action proportional to the $B$-period,
	\begin{equation}
	\label{action-b}
		\mathcal{A} =\alpha\, \partial_t F_0^{(0)}
	\end{equation}
and its associated one-instanton trans-series (\ref{1inst-ex}). The function $T$ is given by (\ref{Tfunction}),
and one obtains (see e.g. the second equation in (\ref{sss})) 
\be
T= -\beta \: \frac{\widehat{E}_2-E_2^D}{12} \: \alpha \: D_t \left(\partial_t F_0^{(0)}\right). 
\ee
In the electric frame, the non-holomorphic $\widehat{E}_2$ becomes simply $E_2$, and with (\ref{tauf0})
	\begin{equation}
		T^e =\alpha, 
	\end{equation}
while the $G$ function becomes
	\begin{equation}
		G^e = \alpha \left( \partial_t F_0^{(0)}(t) + \partial_t \widetilde F^{(0)}(t,\hbar) \right) = \alpha \: \partial_t F^{(0)}(t,\hbar).
	\end{equation}
	With the value $\alpha=1/2$ for the double-well problem, we find 
	\begin{equation}
T^e =\frac{1}{2}, \qquad G^e= \frac{1}{2} t_D(t,\hbar), \qquad  \xi \partial_\lambda G^e = \frac{1}{2} E'.
	\end{equation}
By using (\ref{eq:trans:curlyDerivations}), this also means that
	\begin{equation}
		\mathcal{D}^e = \frac{1}{2} \partial_t.
	\end{equation}
By using these 	results, we can verify that the multi-instanton results obtained in Section \ref{higher-section} reproduce the results obtained from the exact quantization 
condition. Let us take for example the $m=2$ instanton correction, (\ref{eq:trans:2instSolution}). Following the PNP relation, 
the corresponding correction to the energy should be given by
	\begin{equation}
		\partial_\lambda \widetilde F^{(2)} = \left( 
			\frac{\hbar^2 \left(f^{(1)}\right)^2 }{2} \mathcal{D} \partial_\lambda G 
			-\hbar \left(f^{(1)}\right)^2\: \mathcal{D}G \: \partial_\lambda G 
			-2\hbar f^{(2)} \: \partial_\lambda G
		\right) \re^{-2G/\hbar}
	\end{equation}
	and
	\begin{equation}
		E^{(2)} = \xi \partial_\lambda \widetilde F^{(2),e} = \left(  
			\frac{\hbar^2 \left(f^{(1)}\right)^2}{8}E'' 
			 - \frac{\hbar  \left(f^{(1)}\right)^2 }{8} t_D'\:E' -\hbar f^{(2)}\: E'
		\right) \re^{-t_D/\hbar}.
	\end{equation}
	Using values of the constants similar to (\ref{fm-mat}),
		\begin{equation}
		f^{(m)} = \frac{\ri^{m-1}}{m^2 \pi},
	\end{equation}
	we get
	\begin{equation}
		E^{(2)} = \left(  \frac{\hbar^2}{8\pi^2}E''-\frac{\hbar}{8\pi^2} t_D'\:E' - \frac{\ri \hbar}{4\pi} E' \right) \re^{-t_D/\hbar},
	\end{equation}
	precisely what appears in (\ref{eq:exactq:dwEnInstExp}). We have verified the agreement up to $m=5$ (the five instanton correction). 
	
All the results obtained in this section for the double well can be extended to the cubic oscillator studied in \cite{cm-ha}. In that case one has that $\beta=1$,
and the 
relevant instanton action is also of the form (\ref{action-b}) with $\alpha=1$. The function $G$ in the electric frame is also proportional to the quantum $B$-period. 
One can also check that the multi-instanton series obtained in Section \ref{higher-section} reproduce the trans-series obtained from the exact quantization 
condition obtained in e.g. \cite{ddpham,alvarez-casares2}, provided the constants $f^{(m)}$ take the value
\begin{equation}
		f^{(m)} = \frac{\ri^{2m-1}}{2m^2\pi}.
	\end{equation}

\sectiono{Examples of quantum mirror curves: local $\IP^2$}
\label{sec5}

The examples analyzed so far involve Schr\"odinger operators from Quantum Mechanics. We have seen that the trans-series obtained from the refined holomorphic anomaly give 
us new results for the asymptotics of the quantum free energies. These results are compatible with the trans-series obtained with standard techniques 
in Quantum Mechanics. In this section we will consider the spectral problems associated to quantum mirror curves (see \cite{mmrev} for a review and references). In these problems 
there are conjectural exact quantization conditions \cite{ghm,wzh} which determine the spectrum of these operators. This creates the opportunity to compare these exact results with the results obtained with approximation schemes: the all-orders WKB expansion and the 
standard perturbative expansions \cite{hw,gu-s}, as well as their trans-series extensions. 

In this and the next section, we will study the spectral problem for two different toric CY manifolds, local $\IP^2$ and local $\IF_0$, in the all-orders WKB expansion, through the refined holomorphic anomaly equations. We will also calculate the corresponding trans-series. For these spectral problems, there are no trans-series results in Quantum Mechanics to compare with, so the holomorphic anomaly gives the \textit{only} concrete approach to understand their resurgent structure.

\subsection{Refined holomorphic anomaly, trans-series and large-order behavior}

In the case of the local $\IP^2$ geometry, the corresponding quantum-mechanical operator is 
\be
\label{p2-op}
\mO_{\IP^2}=  \re^\mx+ \re^\my+ \re^{-\mx-\my} . 
\ee 
It was conjectured in \cite{ghm} and then proved in \cite{kama,lst} that this operator has a discrete spectrum 
and its inverse $\rho=\mO^{-1}$ is trace class. Although these operators are not of Schr\"odinger type, and they lead to difference equations instead of differential equations, 
one can still use the all-orders WKB approximation \cite{dingle}, as it has been done in \cite{acdkv, hkrs, km,hw}. 
The associated Riemann surface is just the mirror curve of local $\IP^2$, which has the form 
\be
\label{p2curve}
\re^x+ \re^y+ \re^{-x-y} + \kappa=0. 
\ee
The calculation of the classical volume reduces to the calculation of classical periods on this curve. 
We will parametrize the moduli space with the coordinate $z$, which is related to $\kappa$ by 
\be
\label{p2-z}
z={1\over \kappa^3}. 
\ee
The standard, classical periods in the large-radius frame (which is appropriate for the point $z=0$) are given by 
\be 
\label{class-periods}
\ba
-t&=\log(z) + \widetilde \varpi_1(z),\\
{\partial F_0 \over \partial t}&= {1\over 6} \left( \log^2(z) + 2  \widetilde \varpi_1(z) \log(z) +  \widetilde \varpi_2(z) \right), 
\ea
\ee
where
\be
\ba
 \widetilde \varpi_1(z)&= \sum_{j\ge 1} 3 {(3j-1)! \over (j!)^3}  (-z)^j,\\
\widetilde \varpi_2(z)&=\sum_{j \ge 1}{ 18\over j!} 
{  \Gamma( 3j ) 
\over \Gamma (1 + j)^2} \left\{ \psi(3j) - \psi (j+1)\right\}(-z)^{j}. 
\ea
\ee
After integration, we find the prepotential 
\be
\label{p2-prepo}
F_0(t)= {t^3 \over 18}+ 3\re^{-t} -{45 \over 8} \re^{-2t}+ \cdots. 
\ee
Then, a simple calculation shows that (see for example \cite{hk,ghm} for more details)
\be
{\rm vol}_0(E)= 3 {\rd \widehat F_0 \over \rd t} -{\pi^2 \over 2}, 
\ee
where the $\widehat F_0$ symbol means, as in \cite{ghm}, that we changed $\re^{-t} \rightarrow -\re^{-t}$ in the exponentially small corrections appearing in the expansion 
(\ref{p2-prepo}). The relation between $t$ and $E$ is given by 
\be
\label{tE}
t =3 E - \widetilde \varpi_1\left(-\re^{-3 E}\right). 
\ee
Explicitly, one finds \cite{hw}
\be
{\rm vol}_0(E)= {9 E^2 -\pi^2 \over 2} + 9 \sum_{j \ge 1} {(3j-1)! \over j!^3}  \left\{ \psi(3j) - \psi (j+1)-E\right\} \re^{-3j E}. 
\ee
The most efficient way to calculate the higher-order corrections to the quantum volume is to use the refined holomorphic anomaly equations. To set up these equations, we proceed as in e.g. \cite{hkr, hk,hkpk, cesv2}. 
We will use as our global coordinate the modulus $z$ introduced in (\ref{p2-z}). 
We also need some preliminary ingredients from special geometry. We introduce the discriminant of the curve (\ref{p2curve}), 
\be
\Delta=1+27 z, 
\ee
the Yukawa coupling, 
\be
Y=  (\partial_z t)^3 \partial_{t}^3 F_0=-\frac{1}{3} \frac{1}{z^3 \Delta}, 
\ee
and the standard topological string genus-one free energy, 
\be
F_1^{\rm TS}=\frac{1}{2} \log \frac{\partial t}{\partial z} + \frac{1}{12} \log(z^7 \Delta). 
\ee
The holomorphic limit of the propagator $S$, in the large-radius frame, is given by the equation
\be
\partial_z F_1^{\rm TS}= \frac{1}{2} Y S^{\rm LR},  
\ee
where the superscript indicates that $S$ is calculated in the large-radius frame. One finds, explicitly \cite{cesv2}, 
\be
S^{\rm LR}={2\over Y} \left( {1\over z \Delta } - { ~_2F_1 \left( {2\over 3}, {4 \over 3}, 1; -27 z\right)  \over 6 z ~_2F_1 \left( {1\over 3}, {2 \over 3}, 1; -27 z \right)} \right). 
\ee
With these ingredients one can already solve the refined holomorphic anomaly equations in the NS limit, (\ref{has-recursion}). 
The 
initial condition for the recursion is the value of $F_1$ \cite{hk}, 
\be
F_1= -\frac{1}{24} \log(z^{-1} \Delta).
\ee
The holomorphic ambiguity is 
fixed by imposing appropriate boundary conditions. As usual, the holomorphic quantum free energies $F_n$ can be computed in different frames, 
and when needed we will indicate such a frame 
by a superscript. In the conifold frame, and 
near the conifold singularity at $z=-1/27$, the quantum free energies satisfy the gap condition \cite{hk06,hk,hkpk}
\be
\label{p2-con}
F_{n}^C = \frac{(-3)^{n-1} \left(2^{1-2n}-1\right) \left(2n-3\right)!}{\left(2n\right)!} \frac{B_{2n}}{t_c^{2n-2}} + \CO\left(t_c^0\right), \qquad n\ge 2, 
\ee
where $t_c$ is the flat coordinate at the conifold, given by
\be
\label{coni-coord}
t_c = \frac{3 \sqrt{3}}{2\pi} \left(\partial_t F_0-\frac{\pi ^2}{6}\right).  
\ee
Using all this information, it is straightforward to calculate the $F_n$ at very high order in $n$. One finds, for example, 
\be
\ba
F_2 &= Y^2 \left(\frac{S z^4}{128}-\frac{729 z^8}{80}+\frac{27 z^7}{80}-\frac{z^6}{256}\right),\\
F_3 &= -Y^4 \left(\frac{S^3 z^6}{3072}+S^2 \left(\frac{27 z^9}{512}-\frac{z^8}{2048}\right)+S \left(\frac{6561 z^{12}}{640}-\frac{1053
					z^{11}}{5120}+\frac{z^{10}}{4096}\right)- \right.\\ &\left. - \frac{1594323 z^{16}}{1120}+\frac{5137263 z^{15}}{4480}-
					\frac{1644381 z^{14}}{35840}+\frac{2223		z^{13}}{17920}-\frac{z^{12}}{24576}\right).
					\ea
\ee
In order to make contact with the quantum volume, we note that at higher orders in $\hbar$, the relationship between $t$ and $E$ given in (\ref{tE}) gets quantum corrections, 
and one needs the so-called {\it quantum mirror map} $t(E,\hbar)$ \cite{acdkv}. From the point of view of WKB theory, the quantum mirror map just encodes the quantum corrections to the $A$-period, $t$. In this case, the quantum mirror map has the form
\be
\label{p2-qmm}
t(E,\hbar)=3E -3 \left( q^{1/2}+q^{-1/2}\right) \re^{-3E}+ \cdots, \qquad q=\re^{\ri \hbar}. 
\ee
The all-orders perturbative quantum volume is then given by  
\be
\label{p2-vol}
{\rm vol}_{\rm p}(E)= 3 \sum_{n \ge 0} {\partial \widehat F^{\rm LR}_n \over \partial t}\hbar^{2n} -{\pi^2 \over 2}, 
\ee
where the $\widehat F^{\rm LR}_n$, with $n\ge 1$, are obtained by taking the holomorphic large-radius limit of the $F_n$, changing $\re^{-t} \rightarrow -\re^{-t}$ in the exponentially small 
corrections, and then relating $t$ to $E$ via the quantum mirror map (\ref{p2-qmm}).

The first question involving trans-series that we can ask is: what is the large-order asymptotics of the series of quantum free energies 
$ F_n$? General resurgence results predict that the asymptotics should be of form (\ref{lo-as}), 
with the relation (\ref{lo-pred}). Since $z=-\re^{-3E} +\CO(\hbar^2)$ is naturally negative for this problem, 
we will focus on negative values of $z$. A similar problem, concerning the large-order asymptotics of 
the standard topological string free energies of local $\IP^2$, was studied in detail in \cite{cesv2}. The instanton action that 
controls the asymptotic behavior depends on the point where we are in moduli space. 
We will perform the analysis in a region in between the conifold point and the large-radius point, i.e. 
\be
\label{z-region}
{1\over 512}< | z|<{1\over 27}.
 \ee
It turns out that, in this region, the asymptotics of the $F_n$ is controlled by the action 
 \be
 \label{p2-action}
 \CA= {2 \pi \ri \over {\sqrt{3}}} t_c, 
 \ee
 where the conifold coordinate $t_c$ has been defined in (\ref{coni-coord}). This is also the action controlling the 
 asymptotics of the standard topological string free energies in this region, as found in \cite{cesv2}. 
 
 Let us now determine the trans-series associated to this action. We will use the equations (\ref{npNSHAE}) 
to determine the 
 trans-series at the one-instanton level, $F^{(1)}_n$.  We will parametrize the moduli space with the coordinate $z$. 
 We will also need boundary conditions in order to fix the ambiguities. To do this, we proceed as in \cite{cesv2} and 
 we note that, in the conifold frame, we have the behaviour (\ref{p2-con}). By using 
\begin{equation}
		B_{2n} = (-1)^{n-1} \frac{2 (2n)!}{(2\pi)^{2n}} \left(1 + 4^{-n}+\cdots\right)
	\end{equation}
this determines 
\be
b=2
\ee
and the large-order coefficients (\ref{lo-as}) in the conifold frame, 
\be\label{muconcos}
\mu_0^C={1\over 2 \pi^2},  \qquad \mu_n^C=0, \quad n \ge 1. 
\ee
Let us now analyze the equations for the trans-series. First of all, according to (\ref{10eq}), $F_0^{(1)}$ is holomorphic and has no propagator dependence. Therefore, 
this quantity does not depend on the frame and it can evaluated e.g. in the conifold one. By comparing to (\ref{muconcos}), and by using (\ref{lo-pred}), we conclude that
\be
{\Sigma \over 2 \pi \ri}F_0^{(1)}= {1\over 2 \pi^2}. 
\ee
The next correction is non-trivial. By solving the first equation in (\ref{112eq}), we find 
	\begin{equation}
		F_1^{(1)} = f_1^{(1)}(z)  - \partial_z A \: \partial_z F_1^{(0)} \: F_0^{(1)} S, 
	\end{equation}
	where $f_1^{(1)}(z) $ is a holomorphic ambiguity. We fix it again by going to the conifold frame, and by using that $\mu^C_{1} = 0$. Since
	\begin{equation}
		\partial_z F^{(0)}_1 = -\frac{z^2}{8} Y,
	\end{equation}
	we obtain 
	\begin{equation}
		\frac{\Sigma}{2\pi \ri} F_1^{(1)} = \frac{1}{16\pi^2} z^2  \left(\partial_z A\right) Y \: \left(S-S^C\right),
		\label{eq:FI1}
	\end{equation}	
where $S^C$ is the propagator in the conifold frame. It has the explicit expression \cite{cesv2}
	\begin{equation}
	S^C = \frac{z^2}{2} \left(-1-54z+2\:
	\frac{\pi P_{2/3}\left(1+54z\right) +2\sqrt{3} Q_{2/3}\left(1+54z\right)}
	{\pi P_{-1/3}\left(1+54z\right) +2\sqrt{3} Q_{-1/3}\left(1+54z\right)}
	\right), 
	\end{equation}
where $P_\nu(x)$, $Q_\nu(x)$ are Legendre functions. Proceeding in the same way, we solve the second equation in (\ref{112eq}), and we find
\begin{equation}
		\frac{\Sigma}{2\pi \ri} F^{(1)}_2 = \frac{1}{256\pi^2} \Big[z^{2} \: \left(\partial_z A\right)Y
		\left(S-S^C\right)\Big]^2.
	\end{equation}
The pattern we obtain in this way is very similar to what we obtained in the analysis of the Mathieu equation, see e.g. (\ref{Gansatz}). This suggests that the full one-instanton 
amplitude is given by 
\be
\label{p2-f1}
\widetilde F^{(1)} = f^{(1)} \hbar^2 \exp \left[ \frac{\CA + \partial_z \CA \cdot \left(S-S^C\right) \cdot \partial_z \left(F^{(0)}-F^{(0)}_0\right)}{\hbar} \right], 
\ee
with 
\be
{\Sigma \over 2 \pi \ri} f^{(1)}= {1\over 2 \pi^2}. 
\ee
By expanding (\ref{p2-f1}), we reproduce the results obtained above for $F_{1,2}^{(1)}$, and we obtain very explicit expressions for all the coefficients $F_k^{(1)}$. One finds, for example, 
\begin{equation}
\ba
F_3^{(1)}&=\frac{\left(\partial_z A\right) Y^3 z^4  \left(S-S^C\right) }{15360}
			\Big(  5 \left(\partial_z A\right)^2 z^2 \left(S-S^C\right)^2- \\
			& \qquad  - 6 \left(20 S^2+20 S (108 z-1) z^2+\left(209952 z^2-4212 z+5\right) z^4\right)\Big). 
			\ea
			\end{equation}
This leads, through (\ref{lo-pred}), to predictions for all the coefficients $\mu_k$ controlling the large-order behavior of the $F_n$.

	\begin{figure}[h]
		\centering
		\includegraphics[height=4.5cm]{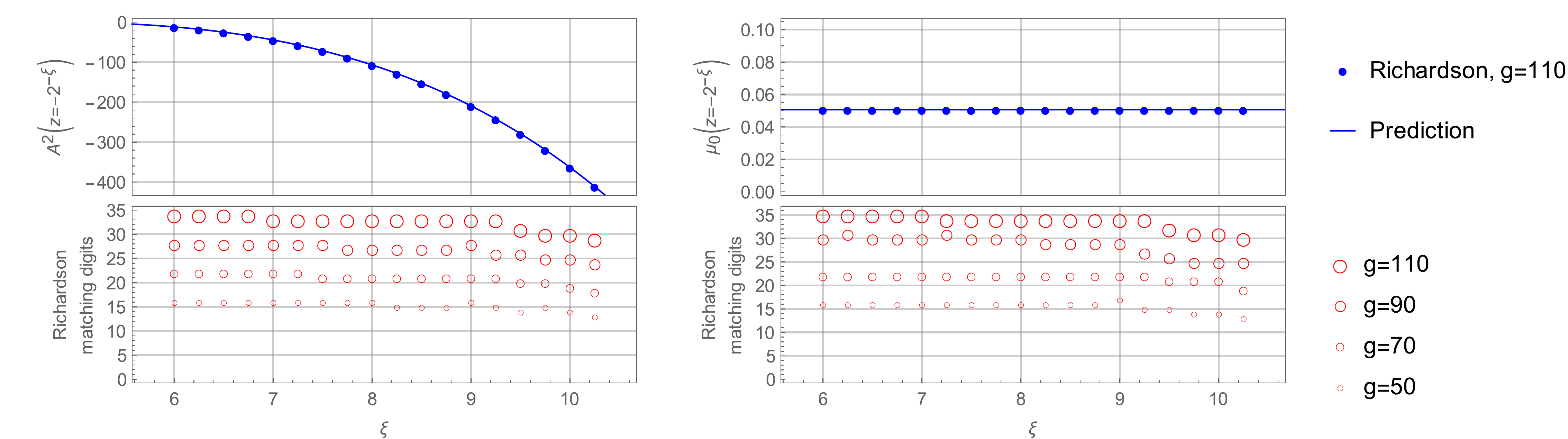}
		\caption{Large-order tests for $\CA^2$ (left) and $\mu_0$ (right), for different values of $z=-2^{-\xi}$.}
\label{fig:largeOrderMod1}
	\end{figure}
\begin{figure}[h]
		\centering
		\includegraphics[height=4.5cm]{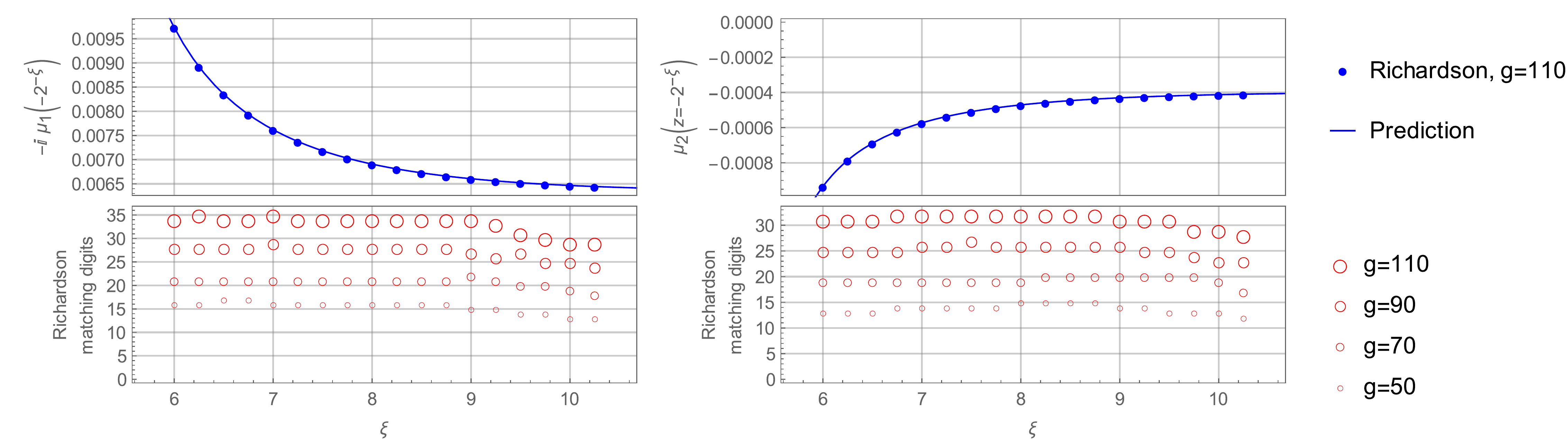}
		\caption{Large-order tests for $\mu_1$ (left) and $\mu_2$ (right), for different values of $z=-2^{-\xi}$.}
		\label{fig:largeOrderMod2}
	\end{figure}
\begin{figure}[h]
		\centering
		\includegraphics[height=4.5cm]{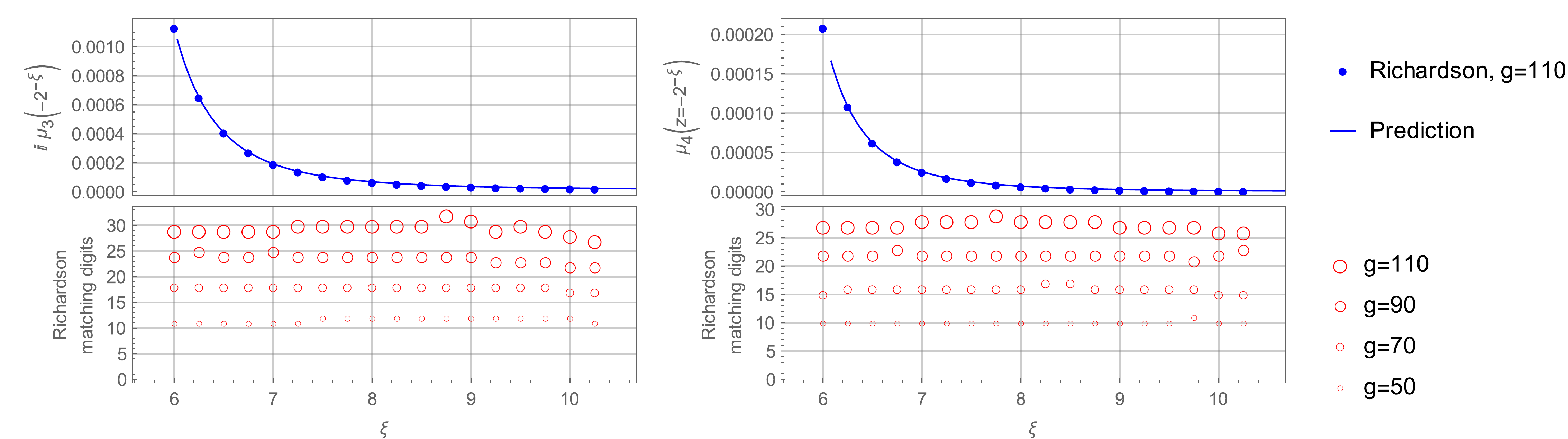}
		\caption{Large-order tests for $\mu_3$ (left) and $\mu_4$ (right), for different values of $z=-2^{-\xi}$.}
		\label{fig:largeOrderMod3}
	\end{figure}

Since we can generate the functions $F_n$ up to a large 
value of $n$, we can test the above expectations in great detail. This is done as follows: we fix a value of the propagator (typically corresponding to a choice 
of frame) and a value of $z$. We use the sequence $F_n(z)$, up to a given value of $n$, to extract numerical approximations for the action $\CA$ and for the 
coefficients $\mu_k$, $k=0,\cdots, 4$, improved with Richardson extrapolation. The numerical results are then compared to the predictions 
coming from (\ref{p2-action}) and (\ref{p2-f1}). 
We show tests of our predictions in \figref{fig:largeOrderMod1}, \figref{fig:largeOrderMod2} and \figref{fig:largeOrderMod3} for $\CA$ and $\mu_0$, for $\mu_{1,2}$, and for $\mu_{3,4}$, respectively. In all cases, we consider the large-radius frame, and values of $z$ of the form $z=-2^{-\xi}$. We indicate the number of matching digits between the numerical approximation and the prediction as a function of the total number of terms in the sequence. As we can see, in the region (\ref{z-region}) the agreement is impressive. However, as we get closer to $z=0$, the number of matching digits decreases. The reason for such a loss of precision was already 
clarified in \cite{cesv2}: near $z=0$ there is another action, given by the large-radius period $t$, which is of the same order than $t_c$, and an 
additional trans-series enters into the asymptotics. We have performed tests for complex values of $z$ and for more general values of the propagator in the region of 
dominance of (\ref{p2-action}), and 
the agreement is again excellent. 
	
\subsection{Quantum free energy: exact versus all-orders WKB}

\begin{figure}[h]
		\centering
		\includegraphics[height=5cm]{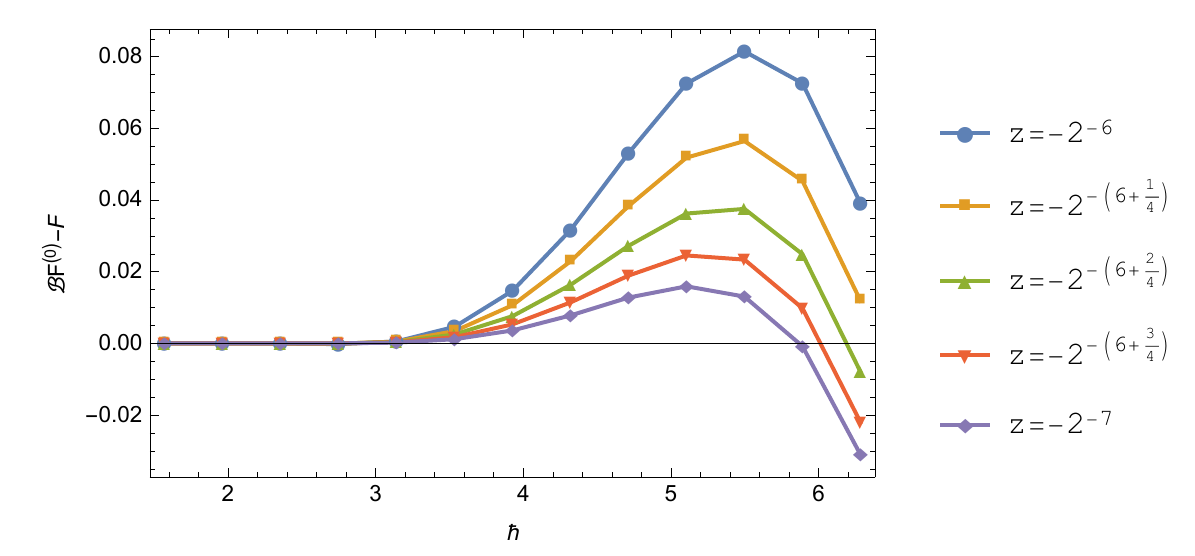}
		\caption{The difference in absolute value between the Borel resummation of the all-orders WKB expansion of the quantum free energy, $\CB F^{(0)}$ and the exact value (\ref{exactF}), as 
		a function of $\hbar$, for different values of $z$.}
		\label{mismatch}
	\end{figure}

As we have already remarked, 
the quantum volume is defined as an asymptotic expansion 
in $\hbar$, and does not always have a non-perturbative definition. The same thing happens with the quantum free energy (\ref{qfe}), which is defined by an asymptotic 
series. It turns out that, in the special case of spectral problems associated to topological strings on toric CY manifolds, there is an exact, 
non-perturbative function whose asymptotic expansion gives back (\ref{p2-vol}). Let us first review how the exact quantum volume is constructed. 
First of all, we note that the quantum free energies $F_n^{LR}(t)$ have an expansion as $t\rightarrow \infty$ of the form 
\be
F_n^{LR} (t)= \delta_{n0} {t^3 \over 18}-\delta_{n1} {t \over 24} + \sum_{k \ge 1} a_{n,k}\re^{-k t}. 
\ee
It turns out that the formal double sum 
\be
F^{LR}_{\rm inst}(t, \hbar) =\sum_{n \ge 0} \sum_{k \ge 1} a_{n,k} \re^{-kt} \hbar^{2n}
\ee
can be first resummed in $\hbar$ in the form \cite{hiv,ikv}
\be
\label{NS-j}
F^{LR}_{\rm inst}(t, \hbar)=\hbar \sum_{j_L, j_R} \sum_{\ell=wd } 
N^{d}_{j_L, j_R}  \frac{\sin\frac{\hbar w}{2}(2j_L+1)\sin\frac{\hbar w}{2}(2j_R+1)}{2 w^2 \sin^3\frac{\hbar w}{2}} \re^{-\ell t}, 
\ee
where $N^{d}_{j_L, j_R}$ are integer numbers, called BPS invariants, which generalize the Gopakumar--Vafa invariants of the CY 
\cite{gv}. This means in particular that the quantum volume can be also resummed in the form 
\be
\label{tp-vol}
{\rm vol}_{\rm p}(E, \hbar) = {t^2 \over 2}- {\pi^2\over 2} -{\hbar^2 \over 8} -3 \hbar 
\sum_{j_L, j_R} \sum_{\ell=wd } 
d N^{d}_{j_L, j_R} (-1)^{wd}  \frac{\sin\frac{\hbar w}{2}(2j_L+1)\sin\frac{\hbar w}{2}(2j_R+1)}{2 w \sin^3\frac{\hbar w}{2}} \re^{-\ell t}. 
\ee
This resummation does not lead to an appropriate quantization condition due to the poles which appear at $\hbar \in 2 \pi \IQ$, 
as first noted in \cite{km}. One needs to add corrections invisible in an $\hbar$ expansion, which were determined in \cite{ghm} as a consequence of a general correspondence between spectral theory (ST) and topological strings (TS), or ST/TS correspondence. The quantization condition was written later in a simpler 
form by using BPS invariants in the NS limit \cite{wzh} (the equivalence between both formulations in many cases was derived in \cite{ggu}.) Let us denote by
\be
f(t, \hbar)= 3 {\partial  \widehat F^{LR}_{\rm inst} \over \partial t} 
\ee
the last term in the r.h.s. of (\ref{tp-vol}). Then, the non-perturbative volume is simply given by 
\be
{\rm vol}_{\rm np}(E, \hbar)= \hbar f \left({2 \pi t \over \hbar}, {4 \pi^2 \over \hbar} \right). 
\ee
The total, exact quantum volume is then defined as 
\be
\label{volex}
{\rm vol}_{\rm ex}(E, \hbar)= {\rm vol}_{\rm p}(E, \hbar) +{\rm vol}_{\rm np}(E, \hbar). 
\ee
There is strong evidence that the above expression defines a function of $E$ and $\hbar$, for real $\hbar$ and $E$ 
sufficiently large, which gives the actual spectrum of the operator (\ref{p2-op}) through the exact quantization condition 
\be
 {\rm vol}_{\rm ex} (E, \hbar)= 2 \pi \hbar \nu. 
 \ee
In some cases, this exact volume function can be derived from a first principles, resummed WKB calculation \cite{szabolcs}. 
It is clear (see e.g. \cite{hatsuda-comments}) that the above procedure also defines an exact function $F^{LR}_{\rm ex}(t, \hbar)$ by 
 \be
 \label{exactF}
 F^{LR}_{\rm ex} (t, \hbar)= {t^3 \over 18}-\hbar^2 {t\over 24} + F^{LR}_{\rm inst}(t, \hbar) + {\hbar \over 2 \pi} F^{LR}_{\rm inst}\left( {2 \pi t \over \hbar}, {4 \pi^2 \over \hbar} \right). 
 \ee
The asymptotic expansion of this function for small $\hbar$ and fixed $t$ is precisely the total quantum free energy (\ref{qfe}) of local $\IP^2$, in the large-radius frame: 
\be
F^{LR}_{\rm ex}(t, \hbar) \sim \sum_{n \ge 0} F_n^{LR}(t) \hbar^{2n}. 
\ee
It is then natural to ask whether the asymptotic series in the r.h.s. is Borel summable, and in case it is, whether its Borel resummation agrees with the exact 
function in the l.h.s. Borel summability in the region of negative $z$ in (\ref{z-region}) follows from the large-order analysis above, 
since $\CA^2$ is negative. We have found numerically that the Borel resummation 
of the $F_n^{LR}$, which we denote by $\CB F^{(0)}$,  {\it differs} from the exact result (\ref{exactF}), as shown in \figref{mismatch}. For example, we obtain
\be
\ba
\CB F^{(0)}\left(  z=-2^{-6}, \hbar=\pi \right)& = 2.0571102\dots,\\
F_{\rm ex} \left(  z=-2^{-6}, \hbar=\pi \right)& = 2.0565565\dots. 
\ea
\ee
Here, $z$ is obtained from $t$ by using the inverse of the classical mirror map given in the first line of (\ref{class-periods}). 
This is in contrast to what happens with the perturbative series in $\hbar$ calculating the energy levels of the spectral problem of (\ref{p2-op}). This series is Borel summable and can be Borel-resummed to the exact values of the energies \cite{hatsuda-comments, gu-s}. At the same time, 
the mismatch we find is not surprising, and it seems to be the default behavior for ``stringy'' series which diverge doubly-factorially, as it has been realized recently in related examples \cite{gmz,cms}. 

Our numerical results suggest that the mismatch between the exact result and the Borel resummation is an exponentially small effect, 
controlled by the same instanton action which was found in \cite{cms}. 
In view of this mismatch, one could ask whether the exact quantum free energy could be recovered by considering the non-trivial trans-series associated to this instanton action and performing Borel--\'Ecalle resummation, as in \cite{cms}. In other words, can we ``semiclassically decode'' the exact function (\ref{exactF}) in terms of its WKB expansion and an appropriate trans-series? 
Without further input, this is a difficult problem, since we have to find the appropriate trans-series parameter, which could depend on both $\hbar$ and the modulus $z$. We leave this issue for future work.

\sectiono{Examples of quantum mirror curves: local $\IF_0$}
\label{sec6}
\subsection{Refined holomorphic anomaly and all-orders WKB}

Let us finally consider another important spectral problem, corresponding to the local $\IF_0$ geometry. The quantum-mechanical operator is 
\be
\label{f0-op}
\mO_{\IF_0}=  \re^\mx+ m_{\IF_0} \re^{-\mx}+\re^\my+ \re^{-\my} . 
\ee 
It has been proved rigorously \cite{kama, lst} that $\mO_{\IF_0}^{-1}$ is trace class and positive when $m_{\IF_0}>0$. 
In this paper we will focus on the special case $m_{\IF_0}=1$, in which the theory simplifies considerably. The all-orders WKB method for 
this operator has been studied in \cite{acdkv, hkrs, km,hw}. The associated Riemann surface is the mirror curve of local $\IF_0$,
\be
\label{F0curve}
\re^x+\re^{-x}+ \re^y+ \re^{-y} + \kappa=0. 
\ee
As in the example of local $\IP^2$, the calculation of the classical volume of phase space reduces to the calculation of classical periods on this curve. 
The appropriate global coordinate in the moduli space is 
\be
z={1\over \kappa^2}. 
\ee
The classical periods are determined by the equations (see e.g. \cite{bt})
\begin{equation}
		\begin{split}
			\partial_z  t& = -\frac{2}{\pi \: z \sqrt{1-16z}} \:\: \mathbf{K}\left(\frac{16z}{16z-1}\right), \\
			\partial_z  \left( \frac{\partial F_0}{\partial t} \right)& = -\frac{2}{z}\:\: \mathbf{K} \left(1-16 z\right),
		\end{split}
		\label{f0-dpers}
	\end{equation}
	where the integration is fixed by the leading-order behavior
	\begin{equation}
	\label{f0-pers}
		\begin{split}
			t &= -\log z - 4 z - 18 z^2 + \cdots,\\
		\partial_t F_0 & = \frac{1}{2}\left(\log z\right)^2 + 4 \left(1 + \log z\right)z + \cdots.
		\end{split}
	\end{equation}
One has in this case that \cite{hk,ghm}
\be
{\rm vol}_0(E)= 2 {\partial F_0 \over \partial t} -{2\pi^2 \over 3},  
\ee
where $t= 2 E+ \CO(\re^{-2E})$ is related to $E$ by the classical mirror map, i.e. by the first equation for the period in (\ref{f0-pers}), once we set $z=\re^{-2E}$. 

As in the case of local $\IP^2$, the higher-order free energies can be easily computed with the NS limit of the 
refined holomorphic anomaly equations. In the special case we are considering with $m_{\IF_0}=1$, one can formulate the problem in terms of 
modular forms, as it was done in \cite{dmp} for the original anomaly equations of \cite{bcov}. We introduce the modular forms, 
	\begin{equation}
		b = \vartheta_2^4\left(q\right), \qquad c = \vartheta_3^4\left(q\right), \qquad  d = \vartheta_4^4\left(q\right),
		\ee
		as well as $\widehat E_2$. The argument $q=\re^{\pi \ri \tau}$ is related to the prepotential $F_0$ by 
	\begin{equation}
		\tau = -{1\over \pi \ri} \frac{\partial^2 F_0}{\partial t^2}, 
	\end{equation}
so that the coefficient in (\ref{tauf0}) is $\beta=-2$, and $Y$ is given by \cite{dmp}
	\begin{equation}
		Y =\frac{2}{d \sqrt{c}} = 2 \: \partial_{ttt} F_0.
	\end{equation}
This provides the map between the modular variables and the geometry of the curve (\ref{F0curve}). In particular, we can recover the modulus as
	\begin{equation}
		z = \frac{1}{16} \frac{b}{c}.
	\end{equation}
 For this model, the NS limit of the refined anomaly equations is of the form (\ref{ha-modular}) with the $\beta=-2$ value. The 
 initial condition for the recursion is $F_1$, which is given by 
\be
F_1=  -\frac{1}{24} \log \left(\frac{1-16z}{z^2} \right)= -\frac{1}{24} \log\left(\frac{256 \, c \, d}{b^2}\right). 
\ee
We now parametrize the holomorphic ambiguity in (\ref{gens-ha}) as 
\begin{equation}
		f_{n,0} = Y^{2n-2}\sum_{i=0}^{3n-3} \alpha_{n,i}\: b^i \: d^{3n-3-i}. 
	\end{equation}
In order to fix the ambiguity, we have to introduce boundary conditions. This requires a discussion of the frames appropriate for different regions in moduli 
space. The large-radius frame, which is appropriate for the region near $z=0$, is defined at the classical level by the 
standard large-radius periods (\ref{f0-pers}). The quantum corrections in this frame are obtained simply by setting $\widehat E_2 \rightarrow E_2(q)$ in the above 
formulae. As explained e.g. in \cite{hkr,dmp,kmz}, there are two other important frames. One is the conifold frame, appropriate near the 
conifold singularity $z=1/16$ (equivalently, near $\kappa=4$). The 
appropriate periods at this point are defined by 
\be
\label{con-periods}
\ba
t_c &= \frac{1}{\pi}\left( \frac{ \partial F_0}{\partial t} -\frac{\pi^2}{3}  \right), \\
\frac{\partial F^C_0}{\partial t_c} &= -\pi \: t + 8 C, 
\ea
\ee
where $C$ is Catalan's constant. The second equation defines the conifold prepotential $F^C_0(t_c)$. The conifold modulus $q_c=\re^{\pi \ri \tau_c}$ is given by 
\be
	\tau_c= -{1\over \pi \ri} \frac{\partial^2F^C_0}{\partial t_c^2} , 
\ee
and it is related to $\tau$ by an $S$ transformation, which can be implemented in the modular forms as 
\begin{equation}
\label{strans}
		b(q) \rightarrow -d \left(q_c\right), \qquad c(q) \rightarrow -c\left(q_c\right), \qquad d(q) \rightarrow -b\left(q_c\right). 
	\end{equation}
The quantum corrections to the free energy in the conifold frame can then be obtained from the non-holomorphic $F_n$ as 
\begin{equation}
		F^C_n = \left(\frac{1}{2}\right)^{2n-2} \left[F_n\right]_{\widehat{E}_2 \mapsto E_2\left(q_c\right)}, 
	\end{equation}
and applying the transformation (\ref{strans}). The gap boundary condition at the conifold reads
\begin{equation}
		F^C_n = \frac{\left(1-2^{1-2n}\right) B_{2n}}{(2n)(2n-1)(2n-2)} \frac{1}{t_c^{2n-2}} + \CO\left(t_c^0\right).
	\end{equation}

To obtain more boundary conditions, we consider the theory in the orbifold 
frame, appropriate near $\kappa=0$. The corresponding periods are given by 
\be
\ba
t_o & = \frac{\ri}{4\pi}\left(2\pi \ri \: t - 2 \: \frac{\partial F_0}{\partial t}\right), \\
\frac{\partial F^O_0}{\partial t_o} &= -\frac{1}{2} \frac{ \partial F_0  }{\partial t}. 
\ea
\ee
 The orbifold modulus $q_o=\re^{\pi \ri \tau_o}$ is given by 
\be
	\tau_o= {1\over \pi \ri} \frac{\partial^2F^O_0}{\partial t_o^2} , 
\ee
and the passage to the orbifold frame is implemented through the modular transformation
\be
\label{lr-or}
	b(q) \rightarrow c(q_o), \qquad d(q) \rightarrow -d(q_o), \qquad c(q) \rightarrow b(q_o).
\ee
The quantum corrections in the orbifold frame can then be obtained from the non-holomorphic $F_n$ as 
\begin{equation}
		F^O_n = \frac{(-1)^{n-1}}{2}\left[ F_n \right]_{\widehat{E}_2 \mapsto E_2(q_o)},
	\end{equation}
	and applying the transformation (\ref{lr-or}). The gap boundary condition at the orbifold is 
\begin{equation}
		F^O_n = \frac{\left(1-2^{1-2n}\right) B_{2n}}{(2n)(2n-1)(2n-2)} \frac{1}{t_o^{2n-2}} + \CO\left(t_o^0\right). 
			\end{equation}
These boundary conditions (together with the absence of constant terms in the expansion at large radius) fix the holomorphic ambiguity completely. One finds, for example, 
\be
F_2 = \frac{ (c+d)^2}{1728 c d^2}\widehat{E}_2 -\frac{37 b^3+51 b^2 d+18 b d^2+20 d^3}{8640 c d^2}.
\ee
In order to obtain the quantum corrections to the quantum volume, we have to take into account that the relation between $t$ and $E$ is now given by the quantum mirror map $t=t(E, \hbar)$, which in this case reads \cite{acdkv}, 
\be
t (E, \hbar)= 2 E-4 \re^{-2 E} -\left( 2 q + 2 q^{-1} +14 \right) \re^{-4 E}+ \CO\left(\re^{-6 E}\right). 
\ee
The all-orders WKB quantization condition is then given by  
\be
\label{f0-allWKB}
{\rm vol}_{\rm p}(E)=2 \sum_{n \ge 0} {\partial F_n \over \partial t} \hbar^{2n}-{2 \pi^2\over 3}=2 \pi \hbar \nu. 
\ee

\subsection{Trans-series and large-order behavior for the energy levels}

We can now use the technology developed in this paper to solve an elementary 
problem in the Quantum Mechanics of the operator (\ref{f0-op}). Let us denote by $\kappa(\nu, \hbar)=\re^{E(\nu,\hbar)}$ the eigenvalue of $\mO_{\IF_0}$, as a function of 
the shifted energy level $\nu=m+1/2$ and $\hbar$. 
One can use standard perturbation theory to find $\kappa(\nu, \hbar)$ a perturbative series in $\hbar$, whose coefficients depend on $\nu$. 
This was first addressed in \cite{hw,hatsuda-comments}, and studied 
more systematically in \cite{gu-s}, who extended the BenderWu package of \cite{bwpack} to include difference equations associated to operators such as (\ref{p2-op}) and (\ref{f0-op}). By using the 
extended package of \cite{gu-s}, one finds, for 
the very first orders, 
\be
\label{pert-kappa}
\kappa(\nu, \hbar)= \sum_{\ell \ge 0} \kappa_{\ell}(\nu) \hbar^\ell= 4 + 2 \nu \hbar + {4 \nu^2 + 1 \over 16}\hbar^2 + {\nu(4 \nu^2+ 3) \over 384} \hbar^3+ \cdots. 
\ee
This result can be in principle derived from the all-orders WKB quantization condition. In analogy with what happened in the Mathieu equation, the quantization condition (\ref{f0-allWKB}) 
defines a quantum dual period 
\be
\label{simwkb}
t_c(\hbar)= {1\over \pi} \left(\sum_{n \ge 0} {\partial F_n \over \partial t} \hbar^{2n}-{ \pi^2\over 3} \right)= \hbar \nu, 
\ee
which is the analogue of $a_D(\hbar)$ in our analysis of the modified Mathieu equation. We can now expand each term $\partial_t F_n$ around 
$\kappa=4$ by using the 
quantum mirror map. This gives
\be
\begin{split}
\hbar \nu &= \frac{\kappa-4 }{2}-\frac{\left(\kappa-4\right) ^2}{32}+\frac{5 \left(\kappa-4\right) ^3}{1536} + \cdots
 +\left(  -\frac{1}{32}+\frac{\kappa-4 }{512}-\frac{5	\left(\kappa-4\right) ^2}{8192}+\cdots \right)\hbar^2+\\
				& + \left(-\frac{13}{49152}+\frac{275 \left(\kappa-4\right) }{1572864} + \cdots \right)\hbar^4 + \cdots.
		\end{split}
	\end{equation}
After inverting this expansion, we obtain, 
\be
\begin{split}
\kappa(\nu, \hbar)&= \left( 4+2 (\hbar \nu) +\frac{(\hbar \nu) ^2}{4}+\frac{(\hbar \nu) ^3}{96}+\cdots\right)
+\left( \frac{1}{16}+\frac{(\hbar \nu) }{128}+\frac{3 (\hbar \nu) ^2}{1024} + \cdots\right)\hbar^2+ \\
			 &+\left(\frac{13}{24576}-\frac{151 (\hbar \nu) }{393216} + \cdots \right)\hbar^4 + \cdots,
		\end{split}
		\label{eq:p1p1:kappaOfHbarNu}
	\end{equation}
which is a rearrangement of the perturbative expansion (\ref{pert-kappa}). 
We now ask the following question: what is the behavior of the coefficients $\kappa_\ell(\nu)$ appearing in 
the perturbative expansion (\ref{pert-kappa}), at large $\ell$ and fixed $\nu$? (this question was asked in e.g. \cite{gu-s}). We expect a behavior of the form 
\begin{equation}
\label{kappalas}
		\kappa_\ell \sim \sum_{r=0}^\infty \frac{\Gamma\left(\ell-b-r\right)}{A^{\ell-b-r}} \mu_r, \qquad \ell \gg 1. 
	\end{equation}
A derivation of this asymptotics directly from the difference equation, via uniform WKB or other techniques, is \textit{not} available. However, we \textit{can} answer this question by using the trans-series for the quantum free energies. By now it should be clear that the one-instanton correction for genus one curves is of the form 
\begin{equation}
\label{f0-1inst}
		\widetilde F^{(1)} = f^{(1)} \hbar^2 \exp \left\{-\frac{\mathcal{A}+ \left(S-\mathcal{S_A}\right)D_t \mathcal{A} \: D_t \widetilde{F}   }{\hbar} \right\},
	\end{equation} 
where the action $\mathcal{A}$ is a period, and $f^{(1)}$ an overall constant. The value of the action can be determined from the large-order behavior of the sequence (\ref{pert-kappa}), and turns out to be
\be
\CA= -2 \pi t = -16 C + 2 {\partial F_0^C \over \partial t_c}. 
\ee
As in the case of the modified Mathieu equation, we have to evaluate (\ref{f0-1inst}) in the magnetic frame. The same argument that led to (\ref{finalG}) produces in here
\begin{equation}
		\begin{split}
			\widetilde F^{(1)}&= f^{(1)} \hbar^2 \: \exp \left\{  -{1\over \hbar} \left(-16 C + 2\frac{\partial F^C_0}{\partial t_c} +
				 2\frac{\partial F^C_1}{\partial t_c} \hbar^2  + \cdots\right) \right\}  \\
				& = f^{(1)} \hbar^2 \:\exp\Big( 2\pi \: t(t_c,\hbar) /\hbar \Big), 
		\end{split}
	\end{equation}
where $t(t_c, \hbar)$ is the quantum version of the second relation in (\ref{con-periods}), i.e. 
\be
\label{tdual}
-\pi t(t_c, \hbar) =\sum_{n \ge 0} {\partial F_n^C \over \partial t_c}\hbar^{2n} + 8 C. 
\ee
We now promote $\nu$ to a trans-series, as in the examples in standard Quantum Mechanics. By using the quantization condition (\ref{simwkb}), we obtain
\be
\Delta \nu^{(1)}= {1 \over \pi \hbar} {\partial \widetilde F^{(1)} \over \partial t} = 2 f^{(1)}\: \re^{16 C/\hbar} \left[\exp \left(-\frac{1}{\hbar}\frac{\partial F^C }{\partial t_c} \right)\right]^{2}.
		\label{eq:p1p1:nuFullOneInstanton}
	\end{equation}
The exponent can be computed explicitly by using our results from the refined holomorphic anomaly, and going to the conifold frame. One finds, 
\begin{equation}
\label{dual-qper}
\ba
			-\frac{1}{\hbar}\frac{\partial F^C }{\partial t_c}  &
				 =\left[\nu  -\nu \log \left(\frac{\nu }{16}\right) +\frac{1}{24 \nu } - \frac{7}{2880 \nu^3} + \CO\left(\nu^{-5} \right) \right]
				\\ & -\nu \log \hbar +\frac{12 \nu ^2+11}{192} \hbar -\frac{20 \nu ^3+49\nu }{4608}\hbar ^2+ \frac{1680 \nu ^4+9240 \nu ^2+889}{2949120}\hbar^3+ \CO\left(\hbar ^4\right).
\ea
	\end{equation}
 A very important property of this result is that the $\hbar$-independent part in the first line (which contains the singularities at $\nu=0$) can be 
exactly resummed in terms of a Gamma function (this happens in all the Quantum Mechanical models analyzed in \cite{cm-ha} and also in the modified Mathieu 
equation, as we saw in (\ref{fnumathieu})). More precisely, it is the large $\nu$ expansion of 
\be
\log\left[ \frac{\sqrt{2\pi} 16^{\nu}}{\Gamma\left(\frac{1}{2}+\nu\right)} \right].
\ee
Putting everything together, we find
	\begin{equation}
		\Delta\nu^{(1)} = f^{(1)}  { 4 \pi \over \Gamma^2 \left(\frac{1}{2}+\nu\right)} \left( {16 \over \hbar} \right)^{2 \nu}  \re^{16 C/\hbar}  \left( 1 + \sum_{k \ge 1} a_k(\nu)  \hbar^k \right), 
		\ee
where the coefficients $a_k(\nu)$ can be easily computed from the  expansion of the functions $\partial_{t_c} F^C_n$. One finds, for the very first coefficients, 
\be
\ba
a_1(\nu)&=\frac{12\nu^2+11}{96}, \\
a_2(\nu)& =\frac{144 \nu ^4-160 \nu ^3+264 \nu ^2-392 \nu +121}{18432}, \\
a_3(\nu)&=   \frac{8640 \nu ^6-28800 \nu ^5+54000 \nu ^4-96960 \nu ^3+188100 \nu ^2-64680 \nu +22657}{26542080}.
\ea
\ee
The trans-series for the energy can be calculated as in \cite{alvarez}, 
\be
\kappa(\nu + \Delta \nu^{(1)}+ \cdots) = \kappa(\nu) + \kappa^{(1)}(\nu)+ \cdots;
\ee
and therefore
\be
\kappa^{(1)}= {\partial \kappa \over \partial \nu}  \Delta \nu^{(1)}= f^{(1)}\, \xi(\nu)\, b_1(\nu), 
\ee
where
\be
\xi(\nu)  ={ 64 \pi \over \Gamma^2 \left(\frac{1}{2}+\nu\right)} \left( {16 \over \hbar} \right)^{2 \nu-1}  \re^{16 C/\hbar} 
\ee
and
\be
b_1(\nu)= 1 + {12 \nu^2 + 24 \nu + 11 \over 96} \hbar +   \frac{144 \nu ^4+416 \nu ^3+552 \nu ^2+136 \nu +193}{18432 }\:\hbar^2+\CO(\hbar^3).
\ee
If we write the one-instanton correction as
\be
\kappa^{(1)} = \hbar^{b} \re^{-A/\hbar} \sum_{r=0}^\infty \kappa^{(1)}_r \hbar^r, 
\ee
standard resurgent analysis predicts that the coefficients in (\ref{kappalas}) are given by 
\be
\mu_r={\kappa^{(1)}_r  \over 2 \pi \ri}. 
\ee
It only remains to determine the coefficient $f^{(1)}$, which can be fixed by the large-order behavior of the sequence $\kappa_\ell(\nu)$ and is given by 
\be
f^{(1)}= {2 \ri \over \pi}. 
\ee
We conclude that the action $A$ and the coefficient $b$ appearing in (\ref{kappalas}) are given by 
\be
A = - 16C,\qquad b = -2\nu+1. 
\ee
The first two coefficients $\mu_0$, $\mu_1$ are given by 
\be
\mu_0 ={4 \over \pi} { 16 ^{2 \nu} \over \Gamma^2 \left(\frac{1}{2}+\nu\right)} , \qquad
\mu_1 = {4 \over \pi} { 16 ^{2 \nu} \over \Gamma^2 \left(\frac{1}{2}+\nu\right)} \frac{12\nu^2+24\nu+11}{96},		
\label{eq:p1p1:largeOrderPredictions}
	\end{equation}
and further coefficients can be computed with the methods explained above. One can extend this trans-series to arbitrary $\nu$ (i.e. not necessarily 
a half-integer) by introducing a factor $(-1)^{2 \nu-1}$ in $f^{(1)}$. 

\begin{figure}[h]\begin{center}
		\includegraphics[height=6cm]{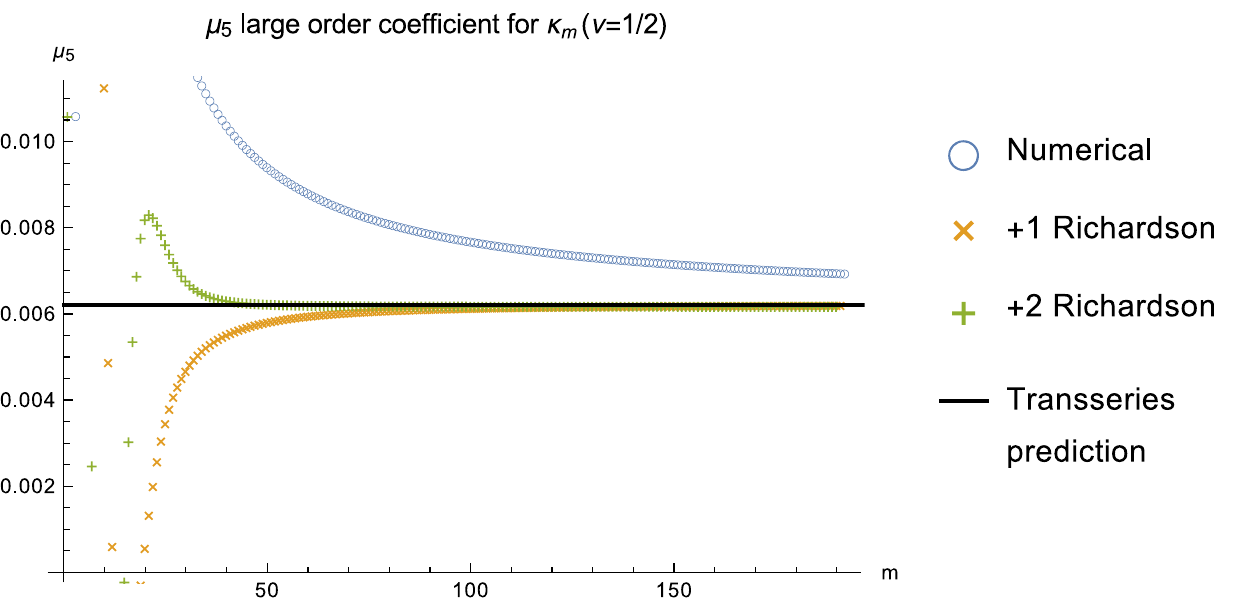}
		\end{center}
		\caption{The sequence (\ref{mui-seq}) for $i=5$, denoted by blue circles, together with its two first Richardson transforms. The convergence to the predicted value (\ref{mu5-pred}) is manifest.}
		\label{fig:p1p1:groundStateLargeOrder}
	\end{figure}

We can test these predictions with an analysis of the sequence $\kappa_\ell(\nu)$ defined in (\ref{pert-kappa}). We extract a numerical prediction $\mu_i^{(k,l)}$ 
for the coefficients $\mu_i$ by taking the first $k$ terms in the sequence 
\be
\label{mui-seq}
\mu_i ^{(m)}=  \left(\frac{m}{A}\right)^i \left(\frac{(-A)^{-b} A^{m}\: \kappa_m}{\Gamma(m-b)}-\sum_{r=0}^{i-1} \frac{\mu_r \: A^r}{(m-b-r)_r}\right),
\ee
which is similar to (\ref{num-mum}), and performing in addition $\ell$ Richardson transforms. As an example, we quote the prediction for the value 
of $\mu_5$ in the ground state $\nu=1/2$, 
\be
\label{mu5-pred}
 \mu_5 = \frac{65131771}{3344302080 \pi } = 0.00619922666015825876905655974809\dots	
 \ee
 while the numerical result by using $120$ terms of the series and $40$ Richardson transforms is 
 \begin{equation}
		\mu_{5}^{(120,40)} = 0.0061992266601582587690566\dots,
	\end{equation}
agreeing on $20$ digits. In \figref{fig:p1p1:groundStateLargeOrder} we plot the sequence (\ref{mui-seq}) for $i=5$, together
	with its first two Richardson transforms. The convergence to the predicted value (\ref{mu5-pred}) is clear. 
	
\sectiono{Conclusions and outlook}
\label{sec7}

In this paper we have extended the correspondence between the refined holomorphic anomaly 
equations and the all orders WKB method into the realm of trans-series, providing 
in this way a \textit{new} method to obtain non-perturbative results in Quantum Mechanics (in one dimension). 
We have constructed trans-series solutions to the NS limit of the refined anomaly equations 
which both recover and generalize known trans-series in standard quantum mechanical models. For spectral problems 
associated to quantum mirror curves, our trans-series solutions 
give information which cannot be obtained by straightforward generalizations of the current methods, 
as we have illustrated in the case of local $\IP^2$ and local $\IF_0$. 

There are clearly many avenues for research open by our new methods. In this paper, our focus 
has been mostly on the one-instanton sector, for which we have in fact produced a universal expression 
(see e.g. (\ref{f0-1inst})). We have also discussed the structure of the higher instanton sectors, and in particular, by 
using PNP relations, we have verified that the results of the 
holomorphic anomaly equations match existing results in Quantum Mechanics. In the case of spectral problems 
associated to quantum mirror curves, the higher instanton sector is less understood. 
One reason is that the PNP relationship in these examples is more 
problematic\footnote{We would like to thank Andrea Brini for detailed discussions on this issue.}. It would be important 
to calculate and test systematically the higher order instanton corrections in the case of quantum mirror curves, and compare them to the exact 
results of \cite{ghm,wzh,cgm}. It has been noticed in \cite{butterfly,hsx, h-bloch} that quantum mirror curves turn out to be related to interesting spectral problems in 
condensed matter systems. It would be interesting 
to see if our methods lead to new non-perturbative results for this type of systems. 

As a tool to analyze instanton trans-series, we have extended the ring of modular forms to take into account exponentially small corrections. This extension requires some unorthodox ingredients from the point of view of the traditional theory of modular forms, but it is very successful in producing correct predictions for the large-order behavior. There might be more natural versions of our formalism, and it would be very interesting to further clarify this new mathematical structure.  

Our comparison of the Borel resummation of quantum free energies and the available exact results is clearly incomplete. First of all, this comparison can be 
done for other examples, such as the modified Mathieu equation, where the exact result for the quantum free energy is provided by instanton calculus \cite{n}. The main question 
is whether all these exact results can be decoded in terms of the perturbative series, plus the trans-series found in this paper. This was achieved in \cite{cms} in a closely 
related example, but a deeper understanding of trans-series parameters is needed in order to have a systematic tool for such an analysis. 

Finally, an important open problem is to extend our discussion (and the one of \cite{cesv1,cesv2}) to the fully refined case, 
involving the two parameters $\epsilon_{1,2}$. This is a necessary step in order to unveil the trans-series structure of the refined topological string, and will hopefully open a new window 
on the non-perturbative structure of the topological string.

\section*{Acknowledgements}
We would like to thank A. Brini and J. Gu for useful discussions. RS would further like to thank CERN TH-Division, and the Kavli Institute for Theoretical Physics, for extended hospitality, where parts of this work were conducted. The work of S.C. and M.M. is supported in part by the Fonds National Suisse, 
subsidies 200021-156995 and 200020-141329, and by the NCCR 51NF40-141869 ``The Mathematics of Physics'' (SwissMAP).  
The research of RS was partially supported by the FCT-Portugal grant UID/MAT/04459/2013. This research was supported in part by the National Science Foundation under Grant No.~NSF~PHY-1125915.

\newpage

\appendix 

\sectiono{Master equation for the refined topological string}
\label{ap-master}
We want to write down a ``master equation'' governing the total free energy of the fully refined topological string. Let 
\be
Z =\exp\left(F_{\rm ref}(t_i; \epsilon_1, \epsilon_2) \right)
\ee
be the total partition function. Then, it can be easily seen that the holomorphic anomaly equations (\ref{pRHAE}) can be written in the form
\be
\label{masterRZ}
\frac{\partial Z}{\partial S} - \frac{1}{2} \epsilon_1 \epsilon_2\, D_z^2 Z = \left( \frac{1}{\epsilon_1 \epsilon_2}\, W + V - U\, D_z \right) Z.
\ee
\noindent
In this equation, the ``boundary functions'' $W$, $V$ and $U$ are given by 
\bea
U &=& D_z F_{(0,0)}^{(0)}, \\
W &=& \sum_{n=0}^{+\infty} \left( \epsilon_1 + \epsilon_2 \right)^{2n} W_n, \\
V &=& \sum_{n=0}^{+\infty} \left( \epsilon_1 + \epsilon_2 \right)^{2n} V_n,
\eea
\noindent
the $\left\{ W_n \right\}$ are given by
\bea
W_0 &=& \frac{\partial F_{(0,0)}^{(0)}}{\partial S} + \frac{1}{2} \left( D_z F_{(0,0)}^{(0)} \right)^2, \\
W_1 &=& \frac{\partial F_{(1,0)}^{(0)}}{\partial S}, \\
W_{n \geq 2} &=& \frac{\partial F_{(n,0)}^{(0)}}{\partial S} - \frac{1}{2} \sum_{m=1}^{n-1} D_z F_{(n-m,0)}^{(0)}\, D_z F_{(m,0)}^{(0)},
\eea
\noindent
and the $\left\{ V_n \right\}$ by
\bea
V_0 &=& \frac{\partial F_{(0,1)}^{(0)}}{\partial S} - \frac{1}{2} \left( D_z^2 F_{(0,0)}^{(0)} \right)^2, \\
V_{n \geq 1} &=& \frac{\partial F_{(n,1)}^{(0)}}{\partial S} - \frac{1}{2} \left( D_z^2 F_{(n,0)}^{(0)} \right)^2 - \sum_{m=1}^{n} D_z F_{(n-m,1)}^{(0)}\, D_z F_{(m,0)}^{(0)}.
\eea

It is not too hard to see that in the standard topological-string limit where $\epsilon_1 \epsilon_2 = g_{\text{s}}^2$, the 
refined master equation \eqref{masterRZ} reduces to the master equation 
obtained in \cite{cesv1}, 
\be
\label{masterZ}
\frac{\partial Z}{\partial S} - \frac{1}{2} g_{\text{s}}^2\, D_z^2 Z = \left( \frac{1}{g_{\text{s}}^2}\, W + V - U\, D_z \right) Z,
\ee
\noindent
and the functions $U$, $V$ and $W$ are given in this limit by
\bea
U &=& D_z F_0^{(0)}, \\
W &=& \frac{\partial F_0^{(0)}}{\partial S} + \frac{1}{2} \left( D_z F_0^{(0)} \right)^2, \\
V &=& \frac{\partial F_1^{(0)}}{\partial S} - \frac{1}{2} \left( D_z^2 F_0^{(0)} \right)^2.
\eea
\noindent
In the NS limit, we recover instead the results discussed in the main text.

\sectiono{Large-order behavior in the modified Mathieu equation}
\label{app-mmathieu}

In this Appendix we give evidence that the trans-series equation for the modified Mathieu equation (\ref{nufnu}) leads to the correct large-order behavior of the perturbative and the 
one-instanton series. The trans-series for the energy reads, 
\be
\label{transen-mathieu}
\ba
E^{(1)}(\nu)&=  {\ri  \over 2 \pi} {\partial  E^{(0)} \over \partial \nu} f(\nu), \\
 E^{(2)}(\nu)&=  {\ri \over 4 \pi} f^2(\nu) {\partial  E^{(0)} \over \partial \nu} -{1\over 8 \pi^2} {\partial \over \partial \nu} \left( f^2(\nu) {\partial E^{(0)} \over \partial \nu} \right). 
 \ea 
\ee
where $f(\nu)$ is given in (\ref{fnumathieu}). We will write, as in \cite{alvarez}, 
\be
f^k(\nu) {\partial  E^{(0)} \over \partial \nu}= \xi^k(\nu) b_k(\nu),  
\ee
where
\be
\xi(\nu)= {2 \pi \over \Gamma^2\left(\nu+{1\over 2}\right)} \left( {32 \over \hbar} \right)^{2 \nu} \re^{{16 \over \hbar}}. 
\ee
One finds, for example, 
\be
\ba
b_1(\nu) &= 1 +{1\over 64} \left(3+ 8 \nu+ 12 \nu^2\right) \hbar+ \cdots, \\
b_2(\nu)&=1+{1\over 32} \left(3+ 4 \nu+ 12 \nu^2\right) \hbar+ \cdots.
\ea
\ee
Let us now focus on the ground state $\nu=1/2$. We write, 
\be
E^{(0)}(1/2)=\sum_{k\ge 0} a_k \hbar^k. 
\ee
Then, the standard dispersion relation 
\be
a_k \sim {1\over \pi} \int_0^\infty {\rm Im}\, E^{(1)}(1/2) {\rd \hbar \over \hbar^{k+1}},
\ee
and the integral
\be
\int_0^\infty \re^{-A/z} z^{-k-2} \rd z= A^{-1-k} \Gamma(1+k),
\ee
give the asymptotic behavior
\be
a_k \sim {2 \over \pi} (-16)^{-k} \left(1 -{5 \over 2k} -{13 \over 8k^2} +\cdots\right), \qquad k \gg 1, 
\ee
which is the result obtained in \cite{stone,dunne-unsal}. However, the trans-series should also give us the asymptotics of the coefficients of the 
first instanton series. To see how this goes, we note that the first instanton correction is 
formally purely imaginary, and it will get an exponentially small real piece related to the real part of the second instanton series. We expect
\be
2 \, {\rm Re} \, E^{(1)}(\nu)=- {1\over 4 \pi^2} {\partial \over \partial \nu} \left( f^2(\nu) {\partial E^{(0)} \over \partial \nu} \right).
\ee
Note the factor of $2$ in the r.h.s., which is standard in resurgence (see e.g. equation (5.14) in \cite{mmnp} or else \cite{as13} for more details). If we write
\be
E^{(1)}(1/2)={32 \ri  \over \hbar} \re^{16/\hbar}\sum_{k \ge 0} a_k^{(1)} \hbar^k,  
\ee
the dispersion relation tells us that
\be
a_k^{(1)}  \sim -{2\over \pi} \int_0^\infty  {\rm Re} \, E^{(1)}_0 (\hbar) {\hbar \over 32}\re^{-16/\hbar} {\rd \hbar \over \hbar^{k+1}}.
\ee
To perform the integral we need the result
\be
\int_0^\infty \re^{-A/z} \log(z) z^{-k-2} \rd z =A^{-k-1} \Gamma (k+1) \left(\log (A)-H_k+\gamma \right),
\ee
where $H_k$ is the harmonic number, as well as the asymptotics at large $k$, 
\be
-H_k+\gamma=-\log(k) -{1\over 2k}+{1\over 12 k^2}+\CO(k^{-3}). 
\ee
One then obtains the prediction for large-order behavior, 
\be
\ba
a_k^{(1)} & \sim -{8 \over \pi} (-16)^{-k} \Gamma(k+1)
\biggl\{ \log(k) \left(1-{4 \over k} +\CO(k^{-2}) \right) \\
 & \, \, \, \, \, \, \, + \log(2) + \gamma + {1\over k} \left( -{3\over 2} - 4 \gamma - 4 \log(2) \right)+ \CO(k^{-2})  \biggr\}, \qquad k \gg 1,
\ea
\ee
which can be checked in detail by using standard techniques\footnote{In the presence of logarithmic terms in the asymptotics, the standard Richardson extrapolation 
method has to be suitably adapted, see for example \cite{zj-expansion,gikm,asv11} for detailed explanations.}. 


\newpage

\bibliographystyle{JHEP}

\linespread{0.6}
\bibliography{biblio}
\end{document}